\documentclass{aa}

\usepackage{graphicx}
\usepackage{amsmath}
\usepackage{amssymb}
\usepackage{xspace}
\usepackage{commath}
\usepackage{times}
\usepackage{bm} 
\usepackage{balance}
\usepackage{hyperref}
\usepackage{mathtools}
\usepackage{stmaryrd}
\usepackage{tabulary}
\usepackage{xcolor}
\usepackage{graphicx}
\usepackage{txfonts}
\usepackage{orcidlink}

\usepackage{afterpage}
\usepackage{multicol}
\newsavebox{\shortpagebox}

\makeatletter
\newcommand{\shortpage}[1]{\par
\setbox\shortpagebox=\vbox{\strut #1\par}\afterpage{\onecolumn
\begin{multicols}{2}
\unvbox\AP@partial
\end{multicols}}\unvbox\shortpagebox
\par}
\makeatother

\hypersetup{dvips, colorlinks=true, linkcolor=blue, citecolor=blue, filecolor=blue, urlcolor=blue}

\defcitealias{Deme+2025}{D25}\defcitealias{Roule+2022}{R22}\defcitealias{Sellwood2015}{S25}

\usepackage{acronym}
\newacro{BL}{Balescu--Lenard}
\newcommand{\BL}{\ac{BL}}

\newcommand{\p}{\partial}
\newcommand{\sgn}{\mathrm{sgn}}
\newcommand{\sech}{\mathrm{sech}}
\newcommand{\deltaD}{\delta_\mathrm{D}}

\newcommand{\ri}{\mathrm{i}}
\newcommand{\re}{\mathrm{e}}

\newcommand{\ra}{\mathrm{a}}

\newcommand{\rd}{{\mathrm{d}}}

\newcommand{\rtot}{\mathrm{tot}}
\newcommand{\DKS}{D_{\mathrm{KS}}}

\newcommand{\mC}{\mathcal{C}}
\newcommand{\mO}{\mathcal{O}}

\newcommand{\mI}{\mathcal{I}}
\newcommand{\mR}{\mathcal{R}}
\newcommand{\mU}{\mathcal{U}}
\newcommand{\mZ}{\mathcal{Z}}
\newcommand{\mM}{\mathcal{M}}

\newcommand{\trel}{t_{\mathrm{rel}}}
\newcommand{\tdyn}{t_{\mathrm{dyn}}}
\newcommand{\Nr}{N_{\mathrm{r}}}
\newcommand{\Mth}{M_{\mathrm{th}}}
\newcommand{\bbR}{\mathbb{R}}
\newcommand{\Ptot}{P_{\mathrm{tot}}}
\newcommand{\cst}{\mathrm{cst}}
\newcommand{\eps}{\epsilon}
\newcommand{\xra}{x_{\mathrm{a}}}
\newcommand{\CDF}{\mathrm{CDF}}
\newcommand{\Etot}{E_{\mathrm{tot}}}
\newcommand{\rhoH}{\rho_{\mathrm{H}}}
\newcommand{\rhoth}{\rho_{\mathrm{th}}}

\newcommand{\scPlummer}{\textsc{Plummer}}
\newcommand{\scHarmonic}{\textsc{Harmonic}}
\newcommand{\scAnharmonic}{\textsc{Anharmonic}}
\newcommand{\scCompact}{\textsc{Compact}}

\newcommand{\bJ}{\mathbf{J}}
\newcommand{\bJp}{\mathbf{J}^{\prime}}
\newcommand{\bk}{\mathbf{k}}
\newcommand{\bkp}{\mathbf{k}^{\prime}}
\newcommand{\bO}{\mathbf{\Omega}}

\newcommand{\oF}{\overline{F}}
\newcommand{\oU}{\overline{U}}
\newcommand{\obT}{\overline{\boldsymbol{\theta}}}
\newcommand{\obJ}{\overline{\mathbf{J}}}
\newcommand{\obJp}{\overline{\mathbf{J}}^{\prime}}
\newcommand{\obk}{\overline{\mathbf{k}}}
\newcommand{\obkp}{\overline{\mathbf{k}}^{\prime}}
\newcommand{\obO}{\overline{\mathbf{\Omega}}}

\usepackage{soul} \usepackage{xcolor}
\definecolor{aquamarine}{rgb}{0.5, 1.0, 0.83}

\begin{document}

\title{Very long-term relaxation \\ of harmonic 1D self-gravitating systems}

\titlerunning{Relaxation of 1D quasi-harmonic systems}

\authorrunning{K. Tep, J.-B. Fouvry \& C. Pichon}

\author{
Kerwann~Tep\inst{1}\orcidlink{0009-0002-8012-4048}
\and
Jean-Baptiste~Fouvry\inst{2}\orcidlink{0000-0002-0030-371X}\and
Christophe~Pichon\inst{2,3}\thanks{Corresponding author: \href{mailto:pichon@iap.fr}{pichon@iap.fr}} \orcidlink{0000-0003-0695-6735}
}

\institute{
Observatoire Astronomique de Strasbourg, CNRS UMR 7550, 11 rue de l’Universit\'e, F-67000 Strasbourg, France 
\and
Institut d'Astrophysique de Paris, CNRS and Sorbonne Universit\'e, UMR 7095, 98 bis Boulevard Arago, F-75014 Paris, France
\and
Kyung Hee University, Dept. of Astronomy \& Space Science, Yongin-shi, Gyeonggi-do 17104, Republic of Korea
}

\abstract{
One-dimensional self-gravitating systems admit genuine thermodynamical equilibria.
For systems with strictly monotonic orbital frequency profile,
the Landau and Balescu--Lenard theories predict a relaxation time scaling linearly with the number of particles, $N$, in agreement with simulations.
Yet, these theories become ill-posed for degenerate frequency profiles,
as is the case in the harmonic potential, where all particles share the exact same mean orbital frequency.
Using an exact collision-driven 1D integrator,
we investigate numerically the self-consistent relaxation of 1D harmonic self-gravitating systems.
We show that harmonic systems relax on a timescale that grows quadratically with $N$.
We show that systems that are only partially degenerate
display the same quadratic scaling for low $N$,
but transition to the linear, non-degenerate behaviour for larger $N$.
The larger the fraction of degenerate orbits,
the larger the value of $N$ at which this transition of dynamical regime occurs.
Finally, we explore the dynamics of fully non-degenerate systems,
albeit with finite radial support:
we confirm that their relaxation time scales linearly with $N$,
though with a substantially larger prefactor than in non-compact systems.
Astrophysically, this investigation should offer some new clues
on the dynamics of density cores, as in the centre of dwarf galaxies.
}


\maketitle

\section{Introduction}
\label{sec:intro}

The dynamics of self-gravitating systems can typically be divided into two phases. First, the system undergoes \textit{violent relaxation}~\citep{LyndenBell1967}, during which its mean potential rapidly evolves toward a more symmetric configuration.
We refer to~\cite{Chavanis2022,Ewart+2022} for detailed reviews on this process of \textit{collisionless relaxation}.
Following this rapid initial phase, the system ends up dynamically frozen
onto some quasi-stationary state, which evolves slowly on long timescales.
In isolated long-range interacting systems,
this latter phase is captured by the inhomogeneous \BL\@~equation~\citep{Heyvaerts2010,Chavanis2012}.
This kinetic equation describes how Poisson shot noise,
originating from the finite number of particles,
drives some long-term relaxation through resonant orbital couplings.
Recently, this kinetic theory has shown remarkable success
in describing the self-consistent relaxation of self-gravitating systems
as diverse as 1D self-gravitating systems~\citep{Roule+2022},
2D razor-thin galactic discs~\citep{Roule+2025}, or 3D globular clusters~\citep{Fouvry+2021},
to name a few.
In all these cases, the system's long-term relaxation
is found to occur on a timescale of order ${ \mO (N \, \tdyn) }$,
with $N$ the total number of particles,
and $\tdyn$ the system's typical dynamical time.

However, the derivation of the Landau and \BL\@ equations relies on several key assumptions. In particular, it assumes that the system is not dynamically degenerate, i.e.\ that the mapping from orbits to their orbital frequencies has a non-zero Jacobian everywhere. In the presence of dynamical degeneracy, these quasilinear kinetic theories no longer apply, and the long-term behaviour of dynamically degenerate systems remains largely unknown.

In this work, we place our focus on such degenerate systems,
whose long-term dynamics we explore numerically.
To reduce the numerical complexity of this work,
we perform two simplifications:
(i) we restrict our analysis to one-dimensional (1D) self-gravitating systems~\cite[see, e.g.\@,][for a review]{Miller+2023};
(ii) we consider, first, the most extreme case of a fully degenerate system,
namely a harmonic system in which all orbits share the exact same frequency.

Astrophysically, the motivations are two-fold.
First, 1D self-gravitating systems are enlightening proxy to model
the formation of cosmological large-scale structures~\cite[see, e.g.\@,][]{Valageas2006,Schulz+2013},
the subtleties of collisionless relaxation~\citep[see, e.g.\@,][]{Joyce+2011,Teles+2011,Colombi+2014},
as well as of its collisional counterpart~\citep[see, e.g.\@,][]{Joyce+2010,Roule+2022},
the small-scale gravitational turbulence~\citep[see, e.g.\@,][]{Nastac+2025,Ginat+2025},
but also the vertical diffusion of stars in galactic discs~\citep[see, e.g.\@,][]{Bennett+2021,Frankel+2023}.

Second, moving from 1D to 3D,
harmonic cores are also key to understand the problem of \textit{core stalling},
namely the slow infall of globular clusters to the central core of dwarf galaxies~\citep[see, e.g.\@,][and references therein]{Inoue2009,Petts+2016,Kaur+2018}.
Indeed, because they are dynamically degenerate,
harmonic cores exhibit a few striking dynamical properties,
such as the failure of the classical formula for dynamical friction~\citep{Read+2006},
some form of super dynamical friction~\citep{Zelnikov+2016},
friction stalling~\citep{Kaur+2022},
and dynamical buoyancy~\citep{Banik+2022}.

This short paper is organized as follows. We describe in Section~\ref{sec:1d_gravity}
the dynamics of 1D self-gravitating systems, and briefly discusses the collision-driven $N$-body integrator.
We present in Section~\ref{sec:relaxation} our method to mesure the relaxation timescales.
Then, we apply it to various potentials of interest.
Finally, we sum up the results in Section~\ref{sec:discussion} and discuss perspective of future projects.
Throughout the main text, technical details are kept to a minimum
and deferred to Appendices or to relevant references.

\section{1D self-gravitating systems}
\label{sec:1d_gravity}

We are interested in the long-term dynamics of one-dimensional self-gravitating systems. From a numerical perspective, such systems are particularly appealing, as their dynamics can be integrated exactly (up to round-off errors) using a collision-driven 
$N$-body integrator~\citep{Noullez+2003}.
This feature allows us to investigate long-term relaxation with tight control over numerical errors.

\subsection{Dynamics}
\label{sec:dynamics_1D}

We consider a 1D system of $N$ particles on an infinite line, with a total mass equal to $M$.
Throughout this paper,
we assume that the cluster comprises particles of equal mass.
On top of their individual kinetic energy,
we suppose that these particles are subject to 1D gravity
described by Poisson equation,
${ \psi'' (x) \!=\! 2 G \rho (x) }$,
with ${ \psi(x) }$ the gravitational potential,
${ \rho (x) }$ the mass density
and $G$ the gravitational constant.
It follows that the gravitational interaction kernel reads
\begin{align}
U(x,x')= G \, |x-x'|.
\label{eq:def_U_1d}
\end{align}
The potential felt by a particule $i$, at location $x_{i}$, is then
\begin{align}
\psi(x_i) = \sum_{j \neq i} m_j \, U(x_i, x_j) = G \sum_{j \neq i} m_j\, |x_i - x_j|,
\label{eq:def_psi_i}
\end{align}
where $m_j$ is the mass of the particle $j$ and $x_j$ its position.
We note that ${ \psi (x_{i}) }$ diverges at infinity,
hence forbidding any escapers. 
From this expression, we can compute the force
${ f_{i} \!=\! - \rd \psi(x_{i}) / \rd x_{i}} $,
imposed on the particle $i$.
It reads
\begin{align}
f_i = G \sum_{j \neq i} m_j \,\sgn(x_i \!-\! x_j)
= G \,\big(M[>\! x_i] \!-\! M[<\! x_i]\big) ,
\label{eq:f_i}
\end{align}
with ${ M [<\! x] }$ (resp.\ ${ M [>\! x] }$)
the total mass enclosed below (resp.\ above) $x$.
We consider a quasi-stationary system of characteristic extent $L$.
Following the virial theorem~\citep[see, e.g.\@,][]{Campa+2014},
its characteristic velocity dispersion is ${ \sigma \!=\! \sqrt{G M L} }$.
As such, the dynamical time is given by ${ \tdyn \!=\! L / \sigma }$.
We let ${ G \!=\! M \!=\! L \!=\! 1 }$ from now on.

\subsection{Time integration}
\label{sec:long_term_relaxation}

We note that the force in equation~\eqref{eq:f_i} is finite
-- though discontinuous --
when the distance between two particles reaches zero.
As such, particle crossing is possible
-- and recurrent -- in 1D systems.
In the absence of any particle crossing,
the force $f_{i}$ is constant.
These two properties are instrumental to construct an exact (up to round-off errors), collision-based integrator.
In practice, we are interested in the (very) long-term relaxation
of small-$N$ systems.
As such, although the integrator is formally exact,
one needs to be careful regarding the accumulation of round-off errors.
We follow the same approach as~\cite{Schulz+2013}
and use a \textit{double float} precision.
This allows us to have particularly well-conserved global invariants
(e.g.\@, relative errors of order ${ 10^{-23} }$, see Figure~\ref{fig:Etot_Ptot}).
This strengthens our confidence in the upcoming numerical explorations.
All these aspects are detailed in Appendix~\ref{app:collision_integrator}.

\subsection{Quasi-stationary equilibria}
\label{sec:models}

Our goal is to explore the long-term relaxation of clusters
with various dynamical properties,
as detailed in Table~\ref{tab:Potentials}.
\begin{table}
\centering
\setlength{\tabcolsep}{.5em}
\begin{tabular*}{0.6711\columnwidth}{@{}|c|c|c|}
\hline
Potential & Compact? & Degenerate?
\\
\hline
\scPlummer{} & No & No
\\
\scCompact{} & Yes & No
\\
\scHarmonic{} & Yes & Yes
\\
\scAnharmonic{} & Yes & Partially
\\
\hline
\end{tabular*}
\caption{Quasi-stationary equilibria considered in the numerical simulations.
These potentials may have a compact (vs. infinite) radial support,
and exhibit some dynamical degeneracy (i.e. particles having
the same mean-field orbital frequencies).
The presence of dynamical degeneracy impacts critically
the long-term relaxation rate of these systems.
We refer to Appendix~\ref{app:QSS} for their detailed definitions.}
\label{tab:Potentials}
\end{table}
First, following~\cite{Roule+2022},
we consider \scPlummer{}:
it is both non-degenerate, i.e.\ all orbits
have initially different orbital frequencies,
as well as with infinite radial support,
i.e.\ the averaged density is non-zero over the whole real axis.
On the contrary, \scCompact{} is still non-degenerate,
but has a compact radial support.
Conversely, \scHarmonic{} is a fully degenerate cluster,
i.e. all particles have the exact same initial orbital frequency.
Additionally, \scHarmonic{} is also of compact support.
This peculiar cluster is the one on which most of the upcoming investigation is focused.
Finally, we also consider \scAnharmonic{}.
It is of compact support, but with a level of dynamical degeneracy
that can be tuned through the parameter $\eps$
(with ${ \eps \!=\! 0 }$ corresponding to \scHarmonic{}).
We refer to Appendix~\ref{app:QSS} for the precise definition
of each of these equilibria.
In practice, we fix units so that, for all potentials,
the total energy is ${ \Etot \!=\! \frac{3}{4} G M^2 L }$ (Appendix~\ref{app:iom}).
As such, on average, all these systems converge towards the same thermodynamical distribution.

\section{Long-term relaxation}
\label{sec:relaxation}

Armed with the potentials from Table~\ref{tab:Potentials},
we can now address the main question
of this work: what is the impact of dynamical degeneracy and/or compact support
on the self-consistent long-term relaxation of self-gravitating systems?

\subsection{Balescu--Lenard equation}
\label{sec:BL}

The long-term evolution of finite-$N$ self-gravitating systems
is typically described by the inhomogeneous \BL\@ equation~\citep{Heyvaerts2010,Chavanis2012}.
In 1D, it takes the form
\begin{align}
\frac{\p F (E,t)}{\p t} \!\propto\! \frac{1}{N} \frac{\p }{\p E} \bigg[ {} & \sum_{k,k'} k \!\! \int\!\! \rd E' \, |U_{kk'}^{\rd}|^{2} \, [...]\,
\deltaD ( k \Omega[E] - k' \Omega[E'] ) \,
\nonumber
\\
\times {} & \, \bigg( k \frac{\p }{\p E} - k' \frac{\p }{\p E'} \bigg) \, F(E,t) \, F(E',t) \bigg] ,
\label{eq:BL}
\end{align}
with ${ F(E,t) }$ the distribution function of the system,
expressed as a function of the individual energy, $E$.
We refer to Appendix~\ref{app:BL_1D}
for the full expression of equation~\eqref{eq:BL}
along with a discussion on its derivation
and its underlying hypotheses.
Equation~\eqref{eq:BL} involves a non-local resonance condition,
through the Dirac delta, ${ \deltaD ( k \Omega[E] - k' \Omega[E'] ) }$,
with ${ \Omega (E) }$ the orbital frequency.
Pairs of particles can interact through the resonance ${ (k,'k') }$,
with an efficiency given by the dressed coupling coefficients,
${ U_{kk'}^{\rd} \!=\! U_{kk'}^{\rd} (E,E',\omega \!=\! k \Omega[E]) }$.
Importantly, equation~\eqref{eq:BL} scales like ${1/N}$.
As such, it describes a relaxation occurring on a timescale
of order ${ N \, \tdyn }$.

For \scHarmonic{},
${ \Omega (E) }$ does not depend on $E$.
As a result, for any resonance ${ k \!=\! k' }$,
this makes the Dirac delta in equation~\eqref{eq:BL} completely ill-defined.
Phrased differently, \BL\@ does not apply to \scHarmonic{},
hence leaving the scaling with $N$ of the relaxation time of \scHarmonic{} essentially unconstrained.
In Appendix~\ref{app:BL_3D},
we emphasise how the same singular behaviour
also occurs in 3D spherically-symmetric systems
with a harmonic mean radial potential.
This is the main motivation for this work:
our goal is to explore numerically the self-consistent relaxation
of fully degenerate systems,
for which the usual (quasilinear) \BL\ kinetic theory does not apply.

\subsection{Measuring relaxation}
\label{sec:methodology}

One-dimensional self-gravitating systems admit a well-defined thermodynamical equilibrium~\citep[see, e.g.\@,][]{Rybicki1971}, toward which all individual realisations ultimately converge.\footnote{This is in stark contrast with 3D self-gravitating,
which, although they continuously relax,
do not admit, stricto sensu, a thermodynamical equilibrium~\citep[see, e.g.\@,][]{Padmanabhan1990,Chavanis+2002,Katz2003}.}
Hence, the level of a relaxation of any given 1D system
can then be assessed through the \textit{distance}
between its instantaneous state
and its thermodynamical equilibrium.

More precisely, let us consider one realisation of a cluster with $N$ particles.
Provided the cluster is recentred around its barycentre,
we can take its total momentum to be zero.
We denote its total energy with $\Etot$.
While $\Etot$ is conserved in time,
it will change from one realisation to the other.
At any given time $t$, we can measure the cumulative mass function for that realisation,
${ M (\leq\! x, t; \Etot, N) }$.
Following Appendix~\ref{app:Thermodynamical_distribution},
this can be compared with its associated thermodynamical equilibrium
${ \Mth (\leq\! x; \Etot , N) }$, via
\begin{align}
\delta M(\leq\! x, t; \Etot , N) = M(\leq\! x, t; \Etot , N) - \Mth (\leq\! x; \Etot , N).
\end{align}
In practice, ${ \delta M }$ is subject to Poisson fluctuations
of order ${ 1 / \sqrt{N} }$.
Ensemble-averaging over $\Nr$ realisations
yields the average difference
\begin{align}
\langle \delta M \rangle (\leq\! x, t) = \big\langle M(\leq\! x, t; \Etot, N) - \Mth (\leq\! x; \Etot , N) \big\rangle .
\label{eq:dM}
\end{align}
By construction, ${ \langle \delta M \rangle }$
is now subject to (much) smaller statistical fluctuations
of order ${ 1 / \sqrt{\Nr \!\times\! N} }$ (see Figure~\ref{fig:thermalization}).
Finally,
we compute the Kolmogorov--Smirnov (KS) distance~\citep{Conover1999}, 
\begin{align}
\DKS (t) = \sup_{x \in \bbR} \, |\langle \delta M \rangle (\leq\! x, t)| .
\label{eq:def_ks}
\end{align}
and define implicitly the effective relaxation time, $\trel$, via
\begin{align}
\DKS (\trel) = D_0,
\label{eq:introduction_D0}
\end{align}
where $D_{0}$ is an ad hoc threshold chosen, with care,
to yield a statistically significant estimate of $\trel$.
It needs to be smaller than the initial value ${ \DKS (t \!=\! 0) }$,
and larger than the noise floor.
For our simulations, 
numerical testing shows that ${ D_0 \!=\! 0.015 }$ and ${ \Nr \!\times\! N \!\sim\! 10^5 }$
offer good compromises.
In Appendix~\ref{app:ks_val_additional},
we check that using the larger ${ D_{0} \!=\! 0.022 }$
does not affect our results.

\begin{figure}
\centering
\includegraphics[width=0.4 \textwidth]{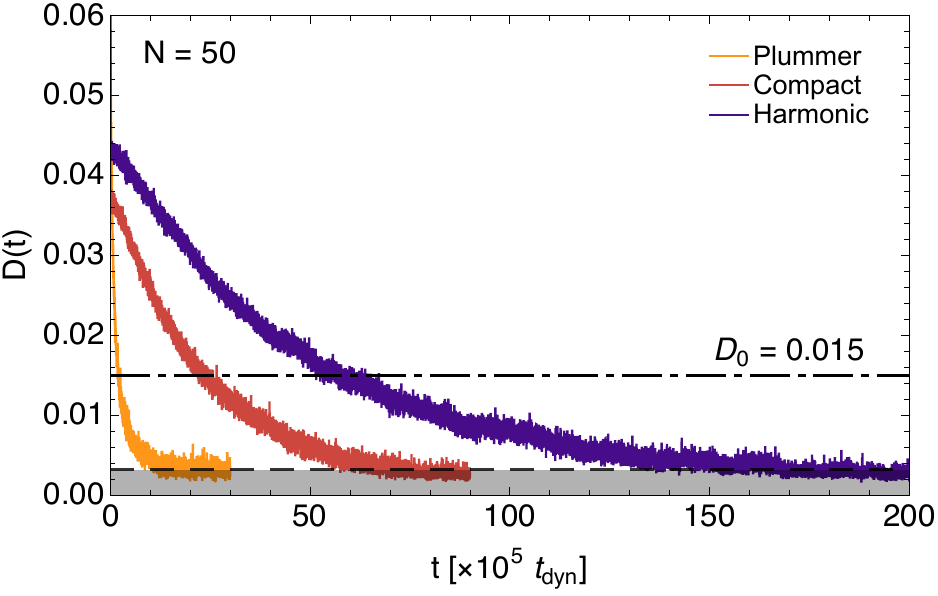}
\caption{Time evolution of the KS distance (equation~\ref{eq:def_ks}) for \scPlummer{}, \scCompact{}, and \scHarmonic{}, with ${ N \!=\! 50 }$ and averaged over ${ \Nr \!=\! 2\, 000 }$ realisations. The bottom dashed line is the characteristic noise level, ${ 1 / \sqrt{\Nr \!\times\! N} }$. The threshold associated with ${ D_{0} \!=\! 0.015 }$ is given by the dot-dashed line.
It is used to estimate the relaxation time, $\trel$ (equation~\ref{eq:introduction_D0}).
The relaxation times vary greatly between the different potentials,
with \scHarmonic{} being the slowest of them all.
}
\label{fig:ks}
\end{figure}

In Figure~\ref{fig:ks},
we illustrate ${ \DKS (t) }$, as a function of time.
In that figure, for the same number of particles 
$N$, and the same average total energy
${ \langle \Etot \rangle }$, we observe a significant variation in the relaxation time among the different potentials, with \scHarmonic{} being the slowest.
From these time series, and following equation~\eqref{eq:introduction_D0}, we determine all the crossing times at which the curve 
${ t \!\mapsto\! \DKS (t) }$ crosses the threshold
$D_0$
between two sampling points.
The relaxation time, $\trel$, is then estimated as the mean of all crossing times, with the error bar indicating the smallest and largest of these values.

\subsection{Application}
\label{sec:application}

\begin{figure}
\centering
\includegraphics[width=0.45 \textwidth]{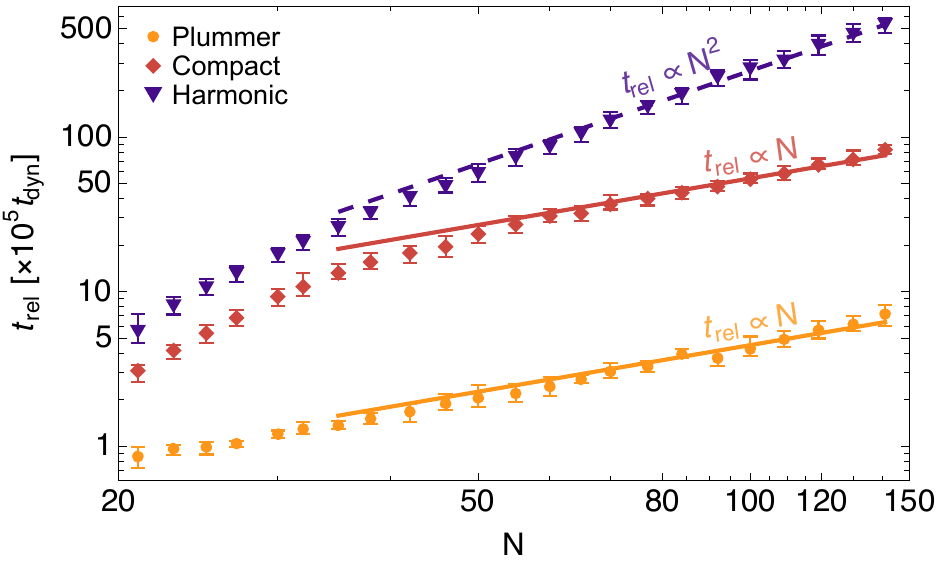}
\caption{Dependence of the relaxation time, $\trel$, with the number of particles, $N$,
for \scPlummer{}, \scCompact{} and \scHarmonic{} (Table~\ref{tab:Potentials}).
Both \scPlummer{} and \scCompact{} relax on a timescale of order ${ \trel \!\propto\! N \tdyn }$ (solid lines),
for $N$ large enough.
On the contrary, \scHarmonic{} relaxes on a timescale of order ${ \trel \!\propto\! N^2 \tdyn }$ (dashed line).
For small $N$, measurements are polluted by small-$N$ effects.
}
\label{fig:scaling_plummer_harmonic}
\end{figure}

Let us now investigate numerically the dependence
of the relaxation time, $\trel$, as a function of the considered potentials.
In practice, we limited ourselves to the range ${ 21 \!\leq\! N \!\leq\! 141 }$,
given the prohibitive number of collisions during relaxation,
of order ${ \mO (N^{2} \trel / \tdyn) }$, see Appendix~\ref{app:collision_integrator}.
Our main result is presented in Figure~\ref{fig:scaling_plummer_harmonic}.

For $N$ large enough, both \scPlummer{} and \scCompact{}
exhibit a relaxation time scaling like ${ \trel \!\propto\! N^{\gamma} \tdyn }$,
where ${ \gamma \!=\! 1.10 \!\pm\! 0.05 }$ for \scPlummer{}
and ${ \gamma \!=\! 1.15 \!\pm\! 0.05 }$ for \scCompact{}.\footnote{The error bars correspond to the 95\% confidence interval over all realisations. For the threshold ${ D_0 \!=\! 0.022 }$, the power law indices read ${ \gamma \!=\! 1.24 \!\pm\! 0.10 }$ for \scPlummer{} and ${ \gamma \!=\! 1.30 \!\pm\! 0.07 }$ for \scCompact{}.}
These values get closer to a linear dependence
as we chose fitting intervals later in time, and lower threshold values $D_{0}$.
This linear dependence in $N$ is fully compatible with \BL\@ (equation~\ref{eq:BL}).

On the contrary, \scHarmonic{} displays a completely different
asymptotic trend in Figure~\ref{fig:scaling_plummer_harmonic}.
Indeed, \scHarmonic{} relaxes on a timescale scaling like
${ \trel \!\propto\! N^{\gamma} \tdyn }$, where ${ \gamma \!=\! 2.19 \!\pm\! 0.05 }$.\footnote{For ${ D_0 \!=\! 0.022 }$,
the power law index reads ${ \gamma \!=\! 2.25 \!\pm\! 0.08 }$.} Taking into account the aforementioned bias, we conjecture a quadratic asymptotic behavior for the relaxation time of \scHarmonic{}.
This is the main result of this paper.
Such an (extremely) slow relaxation cannot be explained by \BL\@,
since this kinetic equation predicts rather
a linear scaling with respect to $N$.
Figure~\ref{fig:scaling_plummer_harmonic} is a rich figure
that deserves further comments.

\subsubsection{Quasi-kinetic blocking}
\label{sec:QKB}

In Figure~\ref{fig:scaling_plummer_harmonic},
we note that the relaxation times of \scPlummer{} and \scCompact{} 
are linear in $N$,
but exhibit a large prefactor, of order $10^{5}$.
This is a phenomenon coined \textit{quasi-kinetic blocking} in~\cite{Roule+2022},
hereafter~\citetalias{Roule+2022}.
We now detail further its origin,
following the same line of reasoning as in~\citetalias{Roule+2022}.

As visible in equation~\eqref{eq:BL},
the relaxations of \scPlummer{} and \scCompact{}
are driven by resonances.
In order to source relaxation at a given $E$,
one must match the orbital frequencies
through the resonance condition
${ k \Omega [E] \!=\! k' \Omega [E'] }$.
Here, ${ (k,k') }$ denotes the resonance at play.
A couple of key remarks are in order
regarding the role played by resonances:
\begin{enumerate}
\item \scPlummer{} and \scCompact{} have a monotonic frequency
profile, ${ E \!\mapsto\! \Omega [E] }$ (see Figure~\ref{fig:Illustration_Frequency}).
As a consequence, any resonance with ${ k \!=\! k' }$
imposes ${ E \!=\! E' }$. Because of the crossed derivative term
in equation~\eqref{eq:BL}, such a resonance drives a vanishing relaxation flux in \BL\@.
\item Because of the system's symmetry,
namely its even density and potential profiles,
the coupling coefficients in equation~\eqref{eq:BL}, $U_{kk'}^{\rd}$,
are non-zero only if $k$ and $k'$ are of same parity~\citepalias[see appendix~{A.4} of][for a detailed justification]{Roule+2022}.
As a result, one must have ${ |k \!-\! k'| \!\geq\! 2 }$
for a given resonance to contribute to the flux of equation~\eqref{eq:BL}.
\item Similarly, because the orbital frequency, ${ \Omega [E] }$, is always positive,
$k$ and $k'$ must have the same sign
to drive relaxation in equation~\eqref{eq:BL}.
\item Introducing $E_{90\%}$ as the typical energy
of the orbit whose apocentre encloses ${\!\sim\! 90\%}$
of the system's total mass,
we find from Figure~\ref{fig:Illustration_Frequency}
that ${  \Omega (E_{90\%}) / \Omega (E \!=\! 0) \!=\! \Omega_{\min} / \Omega_{\max} }$
is ${ \!\sim\! 0.7 }$ for \scPlummer{}
and ${ \!\sim\! 0.8 }$ for \scCompact{}.
Phrased differently, both models sustain
a rather limited range of orbital frequencies.
\item For ${ (k,k') }$ large enough,
as detailed in appendix~{A.4} of~\citetalias{Roule+2022},
one can neglect collective self-gravitating amplification,
so that the dressed coupling coefficients, $U^{\rd}_{kk'}$,
become the bare ones, $U_{kk'}$.
For $k$ large enough, these coefficients exhibit
two types of asymptotic scalings:
for resonance numbers of similar order,
one has ${ U_{kk} \!\sim\! 1/k^{2} }$;
while for resonance numbers of different order,
${ U_{k_{0}k} }$ decreases exponentially with $k$
for fixed $k_{0}$.
\end{enumerate}

Armed with all these remarks, we are now ready to understand
the large prefactor in $\trel$ observed in Figure~\ref{fig:scaling_plummer_harmonic}.
Given the exponential decay of the coupling coefficients
with ${ |k \!-\! k'| }$,
we expect for the main resonance driving the \BL\@ flux
to be of the form ${ (k,k') \!=\! (k, k \!+\! 2) }$,
where we imposed the parity constraint on ${ | k \!-\! k'| }$.
Imposing the resonance ${ k \Omega[E] \!=\! (k \!+\! 2) \Omega[E'] }$
along with the constraint from the limited frequency support,
${ \Omega_{\min} / \Omega_{\max} }$,
we find that the smallest resonance, $k_{\min}$, contributing to the relaxation
is of the order ${ k_{\min} / (k_{\min} \!+\! 2) \!\sim\! \Omega_{\min} / \Omega_{\max} }$.
For \scPlummer{}, we find ${ k_{\min} \!\sim\! 5 }$,
while for \scCompact{}, we find ${ k_{\min} \!\sim\! 8 }$.
Recalling that ${ U_{kk} \!\sim\! 1/k^{2} }$ for large $k$,
we find therefore that the flux predicted by \BL\
is expected to be a factor ${ k_{\min}^4 \!\sim\! 10^{2-3} }$ smaller than one would naively expect.
This decrease is further aggravated by the fact
that the two resonance vectors differ
from one another in the pair ${ (k_{\min} , k_{\min} \!+\! 2) }$,
hence leading to some additional exponential decrease
in the coupling efficiency.
We argue that this combination of effects,
along with a detailed accounting of the exact prefactors in equation~\eqref{eq:BL},
explains the large prefactor, $10^{5}$,
observed for $\trel$ in Figure~\ref{fig:scaling_plummer_harmonic}.
This is the mechanism of quasi-kinetic blocking
put forward in~\citetalias{Roule+2022}.
In a nutshell, it is mainly caused by the fact that only high-order resonances,
hence strongly damped,
can contribute to the \BL\ flux in 1D systems.

Reassuringly, the same line of reasoning
also explains why in Figure~\ref{fig:scaling_plummer_harmonic},
\scCompact{} is found to relax
about ten times more slowly
compared to \scPlummer{}.
As argued before, this is because the narrower frequency range
of \scCompact{} compared to the one of \scPlummer{}
leads to a larger $k_{\min}$ resonance number,
whose contribution to the \BL\ is therefore further reduced.

\subsubsection{Kinetic blockings?}
\label{sec:KB}

We now turn back our interest to the relaxation
of \scHarmonic{} that was observed in Figure~\ref{fig:scaling_plummer_harmonic}
to occur on a timescale of order ${ N^{2} \tdyn }$.
In 1D systems, such a relaxation on ${ N^{2} \tdyn }$ timescales
was already observed in the context of \textit{kinetic blockings}~\citep[see, e.g.\@,][and references therein]{Fouvry2022}.
A kinetic blocking corresponds to a dynamical regime
for which the \BL\@ equation~\eqref{eq:BL} predicts a vanishing flux.
Though, to undergo a kinetic blocking,
the system must satisfy a few assumptions:
\begin{enumerate}
\item This can only occur in 1D.
\item The system can only sustain ${ k \!=\! k' }$ resonances,
for example because of additional symmetries.
\item The frequency profile, ${ E \!\mapsto\! \Omega (E) }$,
has to be strictly monotonic.
\end{enumerate}
If all these three hypotheses are satisfied,
for a given $E$, the resonance condition from equation~\eqref{eq:BL},
namely ${ \deltaD(k \Omega[E] \!-\! k' \Omega[E']) }$,
can only be satisfied for ${ E' \!=\! E }$.
Because of this local resonance,
the crossed term in the second line of equation~\eqref{eq:BL} exactly vanishes.
The two-body driven ${ 1/N }$ relaxation described by \BL\
exactly vanishes: this is a kinetic blocking.
In such systems, the ultimate relaxation towards the thermodynamical equilibrium
can only be driven three-body interactions on the (much) longer ${ N^{2} \tdyn }$ timescale~\citep{Fouvry2022}.

In practice, we argue that no such kinetic blocking
is at play to explain the delayed relaxation of \scHarmonic{},
as visible in Figure~\ref{fig:scaling_plummer_harmonic}.
Indeed, although \scHarmonic{} is 1D [hypothesis 1 above],
it can sustain non-local resonances,
${ k \!\neq\! k' }$.
Indeed, no such constraint
applies to $U_{kk'}^{\rd}$ in equation~\eqref{eq:BL} [hypothesis 2 therefore does not hold].\footnote{This explains, for example,
why \scPlummer{} and \scCompact{} are not subject to a kinetic blocking either.}
But more importantly,
\scHarmonic{} has a constant frequency profile,
in stark constrast with the hypothesis 3 of having a strictly monotonic
frequency profile.
Given these differences,
blocked systems have a vanishing \BL\@ flux in equation~\eqref{eq:BL},
while \scHarmonic{} has a mathematically ill-defined \BL\@ flux.
Phrased differently, in regard to two-body resonances,
blocked systems are \textit{under-resonant},
while \scHarmonic{} is \textit{over-resonant}.
Finally, we point out that the $1/N^{2}$ kinetic equation
that applies to blocked systems \citep[see for example equation~{(4)} in][]{Fouvry2022},
is also ill-defined for the flat frequency profile of \scHarmonic{}.
As such, it does not apply to \scHarmonic{}.
This further strengthens our confidence in the fact
that the delayed relaxation of \scHarmonic{}
is not the same dynamical process as a kinetic blocking.

In practice, \BL\ and its ${1/N^2}$ extensions
are all derived from a quasilinear expansion.
Crucially, such kinetic theories rely on the assumption of phase mixing:
fluctuations, when following their unperturbed mean trajectories,
must shear away from one another as a result of their difference in orbital frequencies.
As a result, on long timescales, only resonant fluctuations,
i.e.\ ones that can stay in phase long enough, can efficiently couple to one another.
This assumption fundamentally breaks down for \scHarmonic{}.
Indeed, within a flat frequency profile,
at leading order, fluctuations do not shear away
from one another in phase space.
At this leading order, perturbations stay in phase,
and this makes traditional perturbative expansions hopeless.
Deriving a closed kinetic equation
to describe the self-consistent relaxation of harmonic systems
falls beyond the scope of this first numerical exploration,
possibly involving techniques stemming from renormalisation theory~\citep{Krommes2002}.

\subsubsection{Thermodynamic blocking}
\label{sec:TB}

In practice, we expect that the delayed relaxation of \scHarmonic{}
is the effective signature of the mechanism of \textit{thermodynamic blocking},
recently put forward in~\cite{Deme+2025}, hereafter~\citetalias{Deme+2025}.
Indeed, placing themselves within the exact same setup as \scHarmonic{},
\citetalias{Deme+2025} provided some analytical insight into the delayed relaxation
of \scHarmonic{},
following a radically different venue compared
to the quasilinear approach of \BL\@.

Starting from the Jeans equations,
i.e.\ the velocity moments of the Vlasov equation~\citep[see, e.g.\@,][]{Binney2008},
\citetalias{Deme+2025} focused on the joint dynamics of the system's mass and kinetic energy densities.
In practice, \scHarmonic{} is continuously subject
to finite-$N$ fluctuations. Because these perturbations are small, the system, 
although perturbed, remains at hydrostatic equilibrium,
i.e.\ remains a quasi-stationary state.
Then, drawing analogies with the gas dynamics and traditional thermodynamics,
\citetalias{Deme+2025} computed the first- and second-order variations
of this system's thermodynamic entropy,
while limiting the physically-allowed perturbations
to the ones that comply with the hydrostratic equilibrium.
In that particular limit, \citetalias{Deme+2025} showed that,
in addition to the expected global Boltzmann-like thermodynamical equilibrium,
there exists a new entropy maximum,
namely harmonic distributions, i.e.\ the present \scHarmonic{} system.
For that system, moving away from dynamical arguments based on resonant couplings,
\citetalias{Deme+2025} showed that the particlar geometry of \scHarmonic{}
prevents macroscopic heat and matter flows,
hence stalling the system's relaxation.
The calculation presented in~\citetalias{Deme+2025}
is a leading order calculation:
we expect therefore that it explains
why no leading order ${1/N}$ relaxation of \scHarmonic{}
is observed in Figure~\ref{fig:scaling_plummer_harmonic}.
Of course, the present arguments remain somewhat qualitative,
and would deserve further scrutiny.
It will be the topic of future work to build upon
the thermodynamical insight from~\citetalias{Deme+2025}
to better characterise the properties
of the slow relaxation of \scHarmonic{},
and the role played by the over-abundance
of resonances in that system.

\subsubsection{Transition in relaxation}
\label{sec:TR}

\begin{figure}
\centering
\includegraphics[width=0.45 \textwidth]{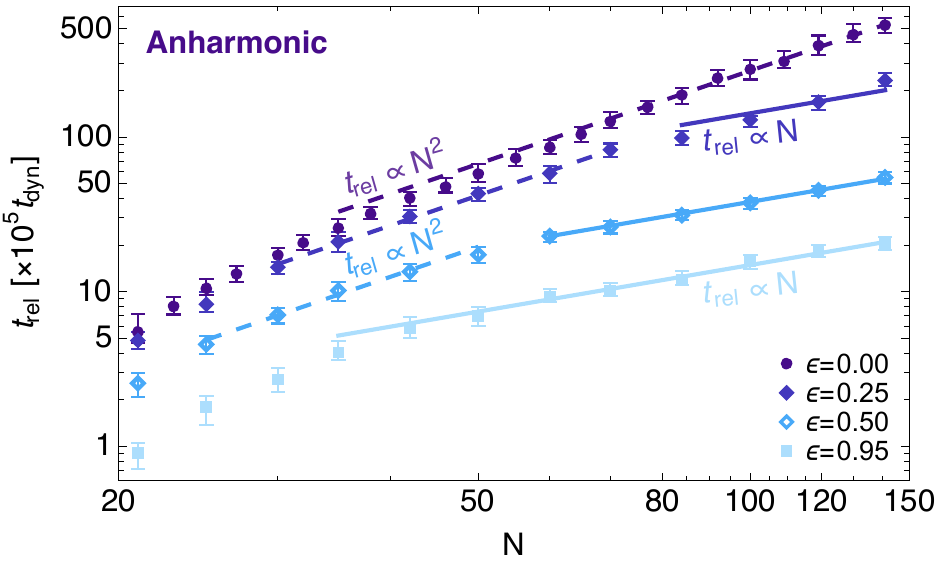}
\caption{Same as Figure~\ref{fig:scaling_plummer_harmonic} for \scAnharmonic{}
and various values of $\eps$, the level of dynamical degeneracy.
As one increases $\eps$, the fraction of non-degenerate orbits,
the clusters relax faster.
Partially degenerate clusters display two relaxation regimes: a relaxation time consistent with ${ N^2 \tdyn }$ (resp.\ ${ N \tdyn }$) for small (resp.\ large) $N$.
Increasing $\eps$ lowers the value of $N$ for which the transition occurs.
}
\label{fig:scaling_anharmonic}
\end{figure}

In order to strengthen our conclusion,
in Figure~\ref{fig:scaling_anharmonic},
we consider \scAnharmonic{} and show the dependence
of its relaxation time with $N$,
as one varies $\eps$, the fraction of non-degenerate orbits
(with ${ \eps \!=\! 0 }$ corresponding to \scHarmonic{}).
First, we note that as one increases $\eps$,
relaxation accelerates.
In addition, partially degenerate clusters exhibit two scaling regimes for their relaxation time:
(i) a scaling like ${ \propto\! N^2 \tdyn }$ for low $N$, just like \scHarmonic{};
(ii) a scaling like ${ \propto\! N \tdyn }$ for large $N$, just like \scPlummer{} and \scCompact{}.
This is clear indication that, indeed, dynamical degeneracy
delays long-term relaxation.
In Figure~\ref{fig:scaling_anharmonic},
we note that the value of $N$ at which the transition occurs increases
as one decreases $\eps$,
i.e.\ as one increases the level of degeneracy of the cluster.
Dynamically, this makes sense.
Indeed, \scAnharmonic{} involves initially ${ N \eps }$ non-degenerate orbits.
At fixed $N$, increasing $\eps$
reduces the number of degenerate orbits,
hence recovering the usual regime of non-degenerate relaxation.
Similarly, at fixed $\eps$, increasing $N$ increases
the number of non-degenerate orbits
up to a point where there are enough of them
to drive the cluster's overall relaxation.

Figures~\ref{fig:scaling_plummer_harmonic} and~\ref{fig:scaling_anharmonic} show jointly how the level of degeneracy directly impacts relaxation timescales.
These two figures only offer a first numerical glimpse
into the impact of dynamical degeneracy on the long-term relaxation
of self-gravitating systems.
A thorough theoretical exploration,
leveraging in particular techniques from \textit{resonance broadening theory}
applied to equation~\eqref{eq:BL}~\cite[see, e.g.,\@][]{Dupree1966,Weinstock1969,Taylor+1971,Dubin2003},
will be the topic of future work.

\section{Conclusions and perspectives}
\label{sec:discussion}

\subsection{Conclusion}
\label{sec:conclusion}

Using an exact collision-driven 1D integrator,
we investigated numerically the self-consistent relaxation of one-dimensional harmonic and quasi-harmonic self-gravitating systems. In particular, in Figures~\ref{fig:scaling_plummer_harmonic} and~\ref{fig:scaling_anharmonic},
we showed that
\begin{itemize}
\item The relaxation time of fully non-degenerate systems,
should they be of infinite or compact radial extent,
scales linearly with $N$.
Nonetheless, relaxation occurs with a substantially larger prefactor
for compact systems.
\item Harmonic systems, because they are dynamically degenerate,
exhibit a relaxation time scaling quadratically with $N$.
\item Systems that are partially degenerate display the same quadratic scaling for low 
$N$, but transition to a linear, non-degenerate behaviour for larger $N$.
\item The larger the fraction of degenerate orbits,
the larger the value of $N$ at which this transition of dynamical regime occurs.
\end{itemize}

\subsection{Perspectives}
\label{sec:perspectives}

This work is only a first step toward understanding
the relaxation of, possibly dynamically degenerate,
long-range interacting systems.
We now conclude by mentioning possible venues for future works,
focusing first on expected impacts in the astrophysical context.

\textit{3D harmonic spheres}. As emphasised in~\cite{Sellwood2015},
the self-consistent relaxation of 3D harmonic globular clusters
is greatly delayed compared to their non-degenerate analogues.
Given its prime astrophysical importance,
a natural next step is to extend the present investigation
to 3D spherical clusters.
This could be made using either direct $N$-body integration codes~\citep[see, e.g.\@,][]{Harfst+2007,Wang+2015,Wang+2020},
or using approximate methods~\citep[see, e.g.\@,][]{Dehnen2014,Mukherjee+2021,Petersen+2025,Tep+2025}
to mitigate the numerical costs.

\textit{Dynamical friction}. As highlighted in the introduction,
harmonic cores are particularly important astrophysically
in the context of the \textit{core stalling} problem,
namely the inefficient sinking of globular clusters and satellites in cored halos~\citep[see, e.g.\@,][and references therein]{Just+2011,DiCintio+2025,Dattathri+2025}.
Here, we focused on self-consistent relaxation,
namely the rate of the change of the system's mean distribution function.
In practice, following a fluctuation-dissipation relation,
this rate of relaxation is typically the combination
of a diffusion and a friction component
(see, e.g.\@, the two terms in equation~\ref{eq:BL}).
As a result, the problem of ``over-resonance'' plaguing
the \BL\ equation in harmonic systems
is just as stringent if one was to infer
dynamical friction from traditional quasilinear kinetic theories like \BL\@.
This particular questioning was one of the prime focus
of the recent work of~\cite{DiCintio+2025}.
Indeed, \cite{DiCintio+2025}
investigated the respective roles of (finite-$N$) fluctuations
and resonances (independent of $N$)
in driving dynamical friction and buoyancy in harmonic cores.
In particular, using tailored numerical simulations,
this work put forward the likely
finite-$N$ origin of core stalling.
Exploring the connexions
between this dynamical signature
and the present delayed relaxation of \scHarmonic{}
is a natural topic for future explorations.

\textit{Core dynamics}. Putting aside the difference in geometry,
we found that harmonic and quasi-harmonic density cores
relax much more slowly than standard kinetic theory predicts~\citep[see, e.g.\@,][]{Inoue2009,Petts+2016,Kaur+2018}.
This affects how long shallow cores in dwarfs can survive and how efficiently they can exchange energy and angular momentum with orbiting substructures.
As such, it could prove useful to interpret dynamically
off-centre active galactic nuclei observed in dwarf galaxies cores,
where standard dynamical friction predicts central coalescence.

\textit{Inexact integration}. All the numerical simulations presented here
were performed using an exact integrator (up to round-off errors).
Yet, its numerical cost tied us to a rather small range of $N$
in Figures~\ref{fig:scaling_plummer_harmonic} and~\ref{fig:scaling_anharmonic}.
One should revisit this analysis for larger $N$,
using a faster, but inexact, integration scheme,
following Appendix~{B} of~\cite{Roule+2022}.
In that case, instantaneous forces are computed
exactly in ${ \mO (N \ln N) }$, through an array sorting.
But motion is integrated approximately using a simple leap-frog scheme.
Given the large errors made by this integrator~\cite[see figure~{10} in][]{Roule+2022},
performing long-term simulations remains surely challenging numerically.

\textit{Multi-mass systems}. Here, we restricted our analysis to a single-mass system.
Following the approach of~\cite{Yawn+1997},
it would be interesting to extend this work to multi-mass systems
and examine the impact on the self-consistent relaxation of harmonic systems.

\textit{Dynamical temperature}. Throughout this work,
we varied the total number of particles, $N$,
hence varying the amplitude of the Poisson shot-noise fluctuations.
Following the approach of~\cite{Fouvry+2023}, it could be interesting instead to vary the system's dynamical temperature, while keeping $N$ fixed.
In practice, this could be achieved by embedding the system within a static background harmonic potential, thereby reducing the level of self-consistency in the fluctuations. Since the dynamics driven by a harmonic potential can be integrated exactly, the present collision-driven integrator could be generalised to this setup while remaining exact, up to round-off errors.

\textit{Thermodynamic blocking}. As discussed in the main text,
\cite{Deme+2025} recently considered the exact same setup
as \scHarmonic{} using arguments based on thermodynamical considerations.
In particular, they put forward how
the constraint of hydrostatic equilibrium makes harmonic distributions
a new entropy maximum, a process they coined \textit{thermodynamic blocking}.
At this stage, one can note that such a thermodynamic argument
greatly differs from any insight that can be gleaned from equation~\eqref{eq:BL}.
One should clarify the connexions between these two point
of views,
as well as predict, a priori, the dependence of the relaxation rate
of \scHarmonic{} with respect to the total number of particles.
This is the topic of current research.

\textit{Time-averaged kinetic theory}. In a harmonic core,
all orbits share the exact same orbital frequency.
As such, one is enticed to describing the system's long-term relaxation
by time-averaging the system's Hamiltonian over this one timescale,
following an approach similar to~\cite{Zelnikov+2016}.
By performing an appropriate canonical transformation, the resulting Hamiltonian closely resembles that of two-dimensional point vortices~\citep[see, e.g.\@,][and references therein]{Chavanis2023}.
Such a rewriting could prove useful to understand
the delay observed in the self-consistent relaxation of harmonic cores (Figure~\ref{fig:scaling_plummer_harmonic}).

\textit{Renormalisation theory}. On long timescales, after averaging over the orbital period, the dynamics of harmonic cores cannot be described using standard quasilinear techniques, as it no longer sustains phase mixing~\citep[see, e.g.\@,][]{Daligault2011}.
As such, it is amenable to renormalisation techniques~\citep[see, e.g.\@,][for a thorough review]{Krommes2002}.
In that context, the \textit{Direct Interaction Approximation}~\citep[see, e.g.\@,][]{Kraichnan1959,Flores+2025}
could offer new clues on self-consistent relaxation.

\textit{2D gravity}. The dynamics of 2D self-gravitating systems
shares deep connexions with the two-dimensional hydrodynamics
of point vortices~\citep[see, e.g.\@,][and references therein]{Chavanis+2007,Bouchet+2012,Chavanis2012Vortex,Chavanis2023}.
In practice, the associated pairwise interaction diverges on small scales~\citep[see, e.g.\@,][]{Fouvry+2025}.
This could affect the efficiency with which harmonic cores
can relax in 2D gravity.
This deserves careful numerical exploration,
using efficient symplectic integration schemes~\cite[see, e.g.\@,][]{Zhang+1993,SanMiguel2006}.

\section*{Data availability}

The data underlying this article 
is available through reasonable request to the authors.
The code \texttt{Gravity1D}, written in {\tt julia}~\citep{JuliaCite}
is available at the URL: \href{https://github.com/KerwannTEP/Gravity1D}{https://github.com/KerwannTEP/Gravity1D}.

\section*{Acknowledgements}

This work is partially supported by the grants ExaSKAle ANR-24-CE31-5182, 
GALBAR ANR-25-CE31-4684 and BEYOND-BL ANR-25-CE57-2626 of the French Agence Nationale de la Recherche.
This project has received financial support from the CNRS through the MITI interdisciplinary programs.
This work has made use of the Infinity Cluster hosted by Institut d'Astrophysique de Paris, partially funded by IDF-DIM-ORIGINES-2023-4-11. 
We thank St\'ephane Rouberol for the smooth running of the
Infinity cluster.

\begin{appendix}

\section{Collision-driven integrator}
\label{app:collision_integrator}

In this Appendix, we detail our implementation
of the exact collision-driven integrator of 1D self-gravitating systems,
following the approach from~\citet{Noullez+2003}.

As argued in equation~\eqref{eq:f_i},
the forces ${ \{ f_i \}_{i} }$ felt by any particles
are constant between two collisions.
As a result, given the position/velocity of particle $i$ at time $t_{a}$,
its motion can be integrated exactly.
For ${ t \!\geq\! t_{a} }$,
as long as it does not undergo any collision,
particle $i$ follows the quadratic motion
\begin{subequations}
\begin{align}
x_i[t] &= x_{i}[t_a] + v_{i}[t_a] (t-t_a) + \tfrac{1}{2} f_i \, (t-t_a)^2.
\label{eq:exact_motion_x}
\\
v_i[t] &= v_{i}[t_a] + f_i \, (t - t_a),
\label{eq:exact_motion_v}
\end{align}
\label{eq:exact_motion}\end{subequations}

Now, let us assume that particle $i$ collides with some particle ${ j \!\geq\! i }$.
In 1D self-gravitating systems,
particles can cross without issue
since the pairwise interaction potential does not diverge for small separation (equation~\ref{eq:def_U_1d}).
Formally, at the time of collision,
we may therefore switch the two particles' indices, velocities and masses, via
\begin{align}
i &\leftrightarrow j \quad;\quad
v_i \leftrightarrow v_j\quad;\quad
m_i \leftrightarrow m_j.
\end{align}
We also need to compute the updated forces after the collision,
following equation~\eqref{eq:f_i}.
Since applying the collision is equivalent to switching indices,
the forces are simply updated through the transformation
\begin{subequations}
\begin{align}
f_i &\rightarrow f_i + G \, (m_i^0 - m_j^0),
\\
f_j &\rightarrow f_j + G \, (m_i^0 - m_j^0).
\end{align}
\label{eq:update_forces}\end{subequations}
Meanwhile, particles that are not involved in the collision
are left unchanged, and so are the forces they are subject to.
From equation~\eqref{eq:update_forces},
we point out that the sum of all the forces felt by the system
increases by ${ 2 (m_i^{0} \!-\! m_j^{0}) }$
after each crossing of the pair of particles ${ (i,j) }$.
In practice, since we only considered systems with equal-mass particles,
this did not require any particular numerical care
to prevent the growth of numerical errors.

Having dealt with one collision, we now need
to determine the time of the subsequent collision,
to finalise the algorithm.
The naive implementation would be to solve the ${ (N \!-\! 1) }$ equations on $t$ that read ${ x_i[t] \!=\! x_{i+1}[t] }$ for ${ i \!=\! 1, ..., N \!-\! 1 }$ (following equation~\ref{eq:exact_motion_x}), and to take the minimum value for $t$.
As pointed out by~\cite{Noullez+2003},
this approach can be drastically accelerated -- from a complexity ${ \mO (N) }$ to ${ \mO (\ln N) }$ --
by implementing a heap structure on the particles, ordering them by increasing collision times.
We refer to~\citet{Noullez+2003} for a detailed description of the heap method and its implementation.

Although this integrator is formally exact,
it accumulates round-off errors at every collision.
These can prove to be a concern for long-term relaxation,
hence requiring additional care.
Indeed, during one dynamical time,
every particle typically crosses every other particle once.
Therefore, the system undergoes ${ \mO (N^{2}) }$ collisions
per dynamical time~\citep[see, e.g.\@,][]{Joyce+2010}.
As shown in~\cite{Roule+2022}, the typical relaxation time for \scPlummer{}
is of order ${ \trel \!\sim\! 10^{5} \tdyn \, N }$,
with the (large) prefactor $10^{5}$
stemming from a \textit{quasi-kinetic blocking} inherent to the 1D geometry.
As such, to reach thermalisation for \scPlummer{},
the integrator must go through a (gigantic) number of collisions,
of order ${ \mO (10^5 \, N^3) }$.
To greatly reduce the impact of round-off error accumulation,
we follow~\citet{Schulz+2013} and use a 80-bit floating-point arithmetic via a \texttt{double float} precision.\footnote{In practice, we use the \texttt{DoubleFloats} library from \texttt{julia} (\href{https://github.com/JuliaMath/DoubleFloats.jl}{https://github.com/JuliaMath/DoubleFloats.jl}). 
It takes advantage of the pre-existing optimisations of \texttt{double} arithmetics.}
This greatly enhances the conservation of the global invariants
(e.g.\@, relative errors of order ${ 10^{-23} }$ instead of ${ 10^{-8} }$),
at the cost of slowing down the code by a factor 4--5.

\begin{figure}
\centering
\includegraphics[width=0.45 \textwidth]{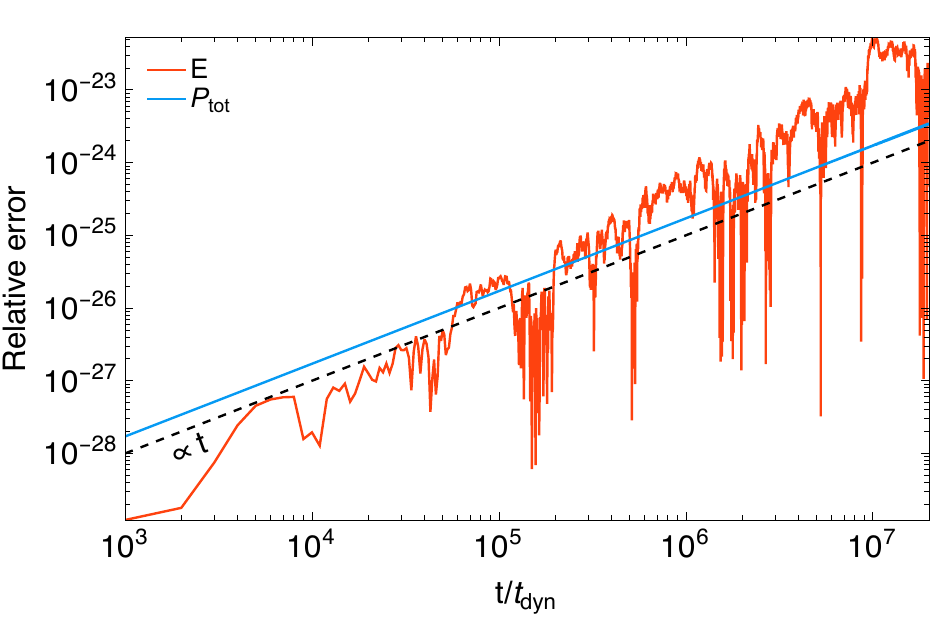}
\caption{Relative error in the total energy (in red) and the total momentum (in blue) during the integration of one \scHarmonic{} cluster with ${ N \!=\! 50 }$,
as considered in Figure~\ref{fig:ks}. 
The errors grow linearly in time (as highlighted by the dashed black line). At ${ t \!=\! 2 \!\times\! 10^7 \tdyn }$,
the integrator went through
${ \!\sim\! 6.5 \!\times\! 10^{9} }$ collisions for that run.
}
\label{fig:Etot_Ptot}
\end{figure}

In Figure~\ref{fig:Etot_Ptot}, we illustrate the performance
of the present scheme,
through the conservation of the total energy (equation~\ref{eq:def_Etot}) and linear momentum (${ \Ptot \!=\! \sum_{i} m_{i} v_{i} }$)
during the (long-time) integration of \scHarmonic{}.
Both invariants exhibit relative errors that are solely driven
by round-off errors.
Errors increase linearly with time,
hinting at some sort of bias in the numerical scheme.
Finally, we point out that the use of the 80-bit floating-point arithmetic via a \textit{double float} precision allows us to keep the energy and momentum error below a (very) satisfactory level (${ \sim\! 10^{-23} }$) over such long timescales.

\begin{figure}
\centering
\includegraphics[width=0.45 \textwidth]{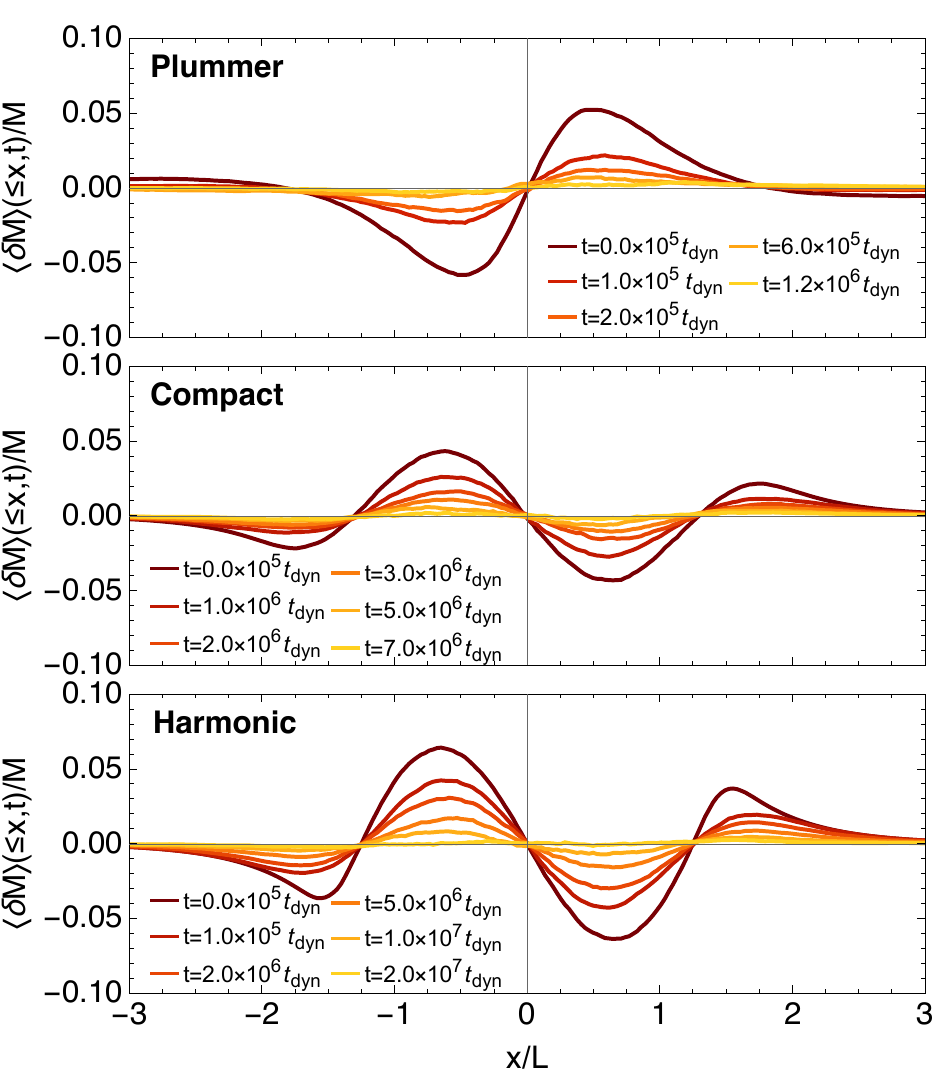}
\caption{Thermalisation of the (ensemble-averaged) \scPlummer{} (top),
\scCompact{} (middle)
and \scHarmonic{} (bottom) clusters used to produce Figure~\ref{fig:ks}.
Each cluster relaxes towards the ${ N \!=\! 50 }$ thermodynamical distribution,
(see Appendix~\ref{app:Thermodynamical_distribution}).
The relaxation timescale greatly differs between the considered potentials. For \scHarmonic{},
the sharp edges of the initial density profile are quickly smoothed out.
}
\label{fig:thermalization}
\end{figure}

In Figure~\ref{fig:thermalization}, we illustrate the difference ${ \langle \delta M \rangle (\leq\! x, t) }$ (equation~\ref{eq:dM}) as a function of time
for \scPlummer{}, \scCompact{} and \scHarmonic{}.
In that figure, we recover their convergence towards the thermodynamical distribution.
We note that each potential takes a different time to thermalize, from ${ \trel \!\sim\! 15 \!\times\! 10^5\, \tdyn }$ for \scPlummer{} up to ${ \trel \!\sim\! 200 \!\times\! 10^5\, \tdyn }$ for \scHarmonic{}.

To obtain Figures~\ref{fig:scaling_plummer_harmonic} and~\ref{fig:scaling_anharmonic},
we performed simulations using $N$ between 21 and 141.
For each value of $N$, we choose the number of realisations, ${ \Nr }$,
so that ${ N \!\times\! \Nr \!\sim\! 100\, 000 }$.
We follow Appendix~\ref{app:ic_generation}
to generate the initial conditions.
For a given realisation, we dumped the position of the particles
every ${ \Delta t \!=\! 1\,000\, \tdyn }$.
In practice, the longest run we performed was for the \scHarmonic{} cluster,
with ${ N \!=\! 141 }$.
Each integration of this cluster until ${t_{\rm rel} }$ required 
${ \sim\! \text{50 h} }$ 
of computation time on a single core.
Such a stringent numerical cost is the main reason
for our use of a rather limited range of $N$ in Figures~\ref{fig:scaling_plummer_harmonic} and~\ref{fig:scaling_anharmonic}.

\section{Integrals of motion}
\label{app:iom}

\subsection{Ensemble-averaged total energy}
\label{app:Ensemble_averaged_energy}

By definition, the exact $N$-body total energy is the sum of the kinetic energy $K$ 
and the potential energy $V$.
It reads
\begin{align}
\Etot {} & = \sum_{i=1}^N \tfrac{1}{2} m_i v_i^2 + G \sum_{i < j}^N m_i m_j \, |x_i \!-\! x_j|
\nonumber
\\
{} & = K + V .
\label{eq:def_Etot}
\end{align}
In the continuum limit, the interation potential becomes
\begin{align}
V= \tfrac{1}{2} \int_{-\infty}^{+\infty} \hspace*{-3mm} \rd x\, \rho(x)\, \psi(x) ,
\end{align}
with ${ \psi (x) }$ and ${ \rho(x) }$, the mean-field potential and density.
Applying the virial theorem for 1D self-gravitating clusters~\citep[see, e.g.\@,][]{Campa+2014},
namely ${ 2 K \!-\! V \!=\! 0 }$,
then yields
\begin{align}
\Etot = \tfrac{3}{4} \int_{-\infty}^{+\infty} \hspace*{-3mm} \rd x \, \rho(x) \, \psi(x).
\label{eq:calc_Etot}
\end{align}

\subsection{Orbital frequency}
\label{app:action}

Let us consider a particle with energy $E$ evolving
in a fixed symmetric 1D potential, $\psi$.
Its specific energy is
\begin{align}
E = \frac{1}{2} \dot{x}^2 + \psi(x).
\end{align}
We can rearrange this equation and obtain
\begin{align}
\dot{x} =\pm \sqrt{2(E \!-\! \psi[x])}.
\end{align}
It follows that the orbital period of the particle, $T$, reads
\begin{align}
\frac{T}{2} = \int_{0}^{T/2}\hspace{-4mm} \rd t = \int_{- \xra}^{\xra} \frac{\rd x}{|\dot{x}|}
= \int_{- \xra}^{\xra} \hspace{-2mm}\frac{\rd x}{\sqrt{2(E \!-\! \psi[x])}},
\end{align}
where ${ \xra \!>\! 0 }$ is the turning radius of the orbit,
i.e.\ it satisfies ${ \psi (\xra) \!=\! E }$. 
Since the orbital frequency $\Omega$ satisfies the relation ${ \Omega T \!=\! 2\pi }$,
we obtain
\begin{equation}
\frac{1}{\Omega} = \frac{1}{\pi} \int_{- \xra}^{\xra} \hspace{-2mm} \frac{\rd x}{\sqrt{2(E \!-\! \psi[x])}}
= \frac{\sqrt{2}}{\pi} \int_{0}^{\xra} \hspace{-2mm} \frac{\rd x}{\sqrt{E \!-\! \psi(x)}} .
\label{eq:Omega}
\end{equation}

\section{Quasi-stationary equilibria}
\label{app:QSS}

In this Appendix, we detail the construction of the various
quasi-stationary equilibria considered in the main text.
Fortunately, in 1D, this can be made easily using Eddington inversion,
as we first recall.

\subsection{Eddington inversion}
\label{app:eddington}

Applying equation~{(B.74)} of~\citet{Binney2008},
the DF of a 1D system reads
\begin{equation}
F(E) =-\frac{1}{\pi\sqrt{2}} \int_E^{\infty} \hspace*{-3mm} \frac{\rd \Psi}{ \sqrt{\Psi \!-\! E}} \,\frac{\rd \rho}{\rd \psi}(x[\Psi]),
\end{equation}
where ${ x [\Psi] }$ is solution of ${ \Psi \!=\! \psi(x) }$ and ${ \rd \rho / \rd \psi \!=\! \rho' / \psi' }$.
Changing the integration variable from $\Psi$ to $x$ yields
\begin{equation}
F(E) =-\frac{1}{\pi\sqrt{2}} \int_{x_{\ra}}^{\infty} \hspace*{-3mm} \frac{\rd x\,\rho'(x)}{ \sqrt{\psi(x) \!-\! E}} ,
\label{eq:final_Eddington}
\end{equation}
where $x_{\ra}$ is the boundary of motion,
satisfying ${ E \!=\! \psi (x_{\ra}) }$. We shall also use the notation ${ F(x_{\ra}) \!=\! F (x_{\ra} [E]) }$ to refer to this DF.

Using equation~\eqref{eq:final_Eddington},
we are now set to compute the DF of the clusters considered in the main text.
In practice, to ease the comparison between the different clusters,
following Appendix~\ref{app:Ensemble_averaged_energy},
for all of them, we set their total energy to ${ \frac{3}{4} G M^2 L }$,
with $M$ the cluster's total mass,
and $L$ its characteristic length.

\subsection{\scPlummer{}}
\label{app:Plummer}

\begin{figure}
\centering
\includegraphics[width=0.45 \textwidth]{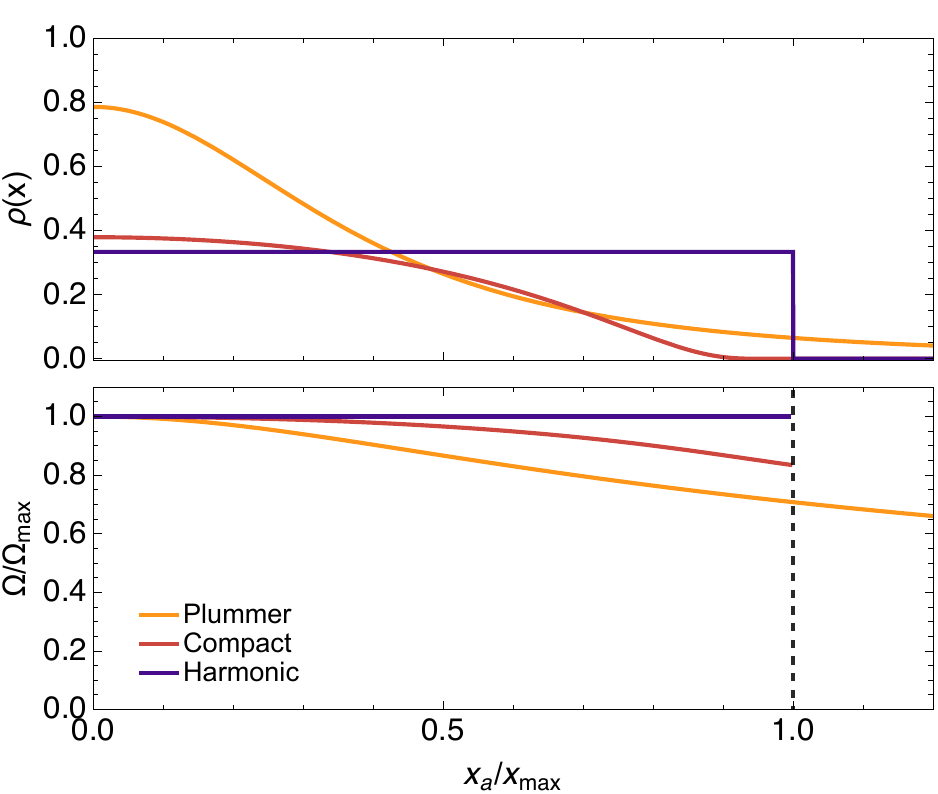}
\caption{Illustration of the density (top panel) and frequency (bottom panel) profiles
associated with \scPlummer{} (Appendix~\ref{app:Plummer}),
\scCompact{} (Appendix~\ref{app:Compact})
and \scHarmonic{} (Appendix~\ref{app:Harmonic}). We define $\Omega_{\max}$ as the maximum frequency of each system.
We also introduce $x_{\max}$ as the maximum populated pericentre in compact systems, and as ${ x_{\max} \!=\! x_{90\%} }$ for \scPlummer{},
such that ${ [-x_{90\%},x_{90\%}] }$ contains 90 \% of the total mass.
The families differ in their finite/infinite radial support,
and the presence/absence of dynamical degeneracies.
}
\label{fig:Illustration_Frequency}
\end{figure}

We first consider the \scPlummer{} potential.
Following~\cite{Roule+2022},
it reads
\begin{subequations}
\begin{align}
\rho(x) & = \frac{M}{2\alpha} [1 \!+\! (x/\alpha)^2]^{-3/2} ,
\\
\psi(x) & = E_{\alpha} \sqrt{1 \!+\! (x/\alpha)^2},
\\
F(E) & = \frac{15 M }{32\!\sqrt{2} \, \sigma_{\alpha}\alpha}\, (E/E_{\alpha})^{-7/2},
\end{align}
\label{eq:def_Plummer}\end{subequations}
where ${ \alpha \!=\! 2L / \pi }$.
We also introduced ${ \sigma_{\alpha} \!=\! \sqrt{G M \alpha} }$,
${ E_{\alpha} \!=\! G M \alpha }$.
The total energy is ${ \Etot \!=\! \frac{3\pi}{8} G M^2 \alpha \!=\! \frac{3}{4} G M^2 L }$.
In Figure~\ref{fig:Illustration_Frequency}, 
we illustrate the associated density.

\subsection{\scCompact{}}
\label{app:Compact}

We define the \scCompact{} density by letting
\begin{align}
\rho(x) &= \frac{M}{ \mC a}
\begin{cases}
\re^{- a^{2} / (a^{2} - x^{2})} & \mathrm{if} \, |x|< a ,
\\
\hspace*{10mm} 0 & \mathrm{otherwise},
\end{cases}
\label{eq:def_rho_anharmonic}
\end{align}
with ${ \mC \!=\! \int_{-1}^{1} \rd y \,\re^{- 1 / (1 - y^{2})} \!\simeq\! 0.443994 }$.
This corresponds to a compact system,
i.e.\ one with a finite radial support
within the domain ${ |x| \!\leq\! a }$.
Following equation~\eqref{eq:def_psi_i},
the associated potential reads
\begin{align}
\psi(x) & \!=\! GM 
\begin{cases}
\frac{a}{\mC}\int_{-1}^{1} \rd y\, |y \!-\! x/a| \, \re^{- 1 / (1 - y^{2})} & \mathrm{if} \, |x|< a ,
\\
\hspace*{15mm} |x| & \mathrm{otherwise}.
\end{cases}
\end{align}
Using equation~\eqref{eq:final_Eddington},
we can compute the DF.
If ${ x_{\ra} \!\geq\! a }$, then ${ F(x_{\ra}) }$ vanishes.
Otherwise, one finds
\begin{equation}
F(x_{\ra}) =\frac{M \sqrt{2}}{\mC a^2 \pi} \int_{x_{\ra}}^{a} \frac{\rd x\, (x/a)}{[1 \!-\! (x/a)^2]^2} \frac{\re^{- a^{2} / (a^{2} - x^{2})}}{\sqrt{\psi(x) \!-\! \psi(x_{\ra})}}.
\end{equation}
Finally, we find numerically that the total energy is
${ \Etot \!\simeq\! 0.343104 \, G M^2 a \!\simeq\! \frac{3}{4} \,G M^2 L }$,
with ${ a \!\simeq\! 2.18593\, L }$.

\subsection{\scHarmonic{}}
\label{app:Harmonic}

We now focus on the \scHarmonic{} potential.
We consider a system with constant density
within the interval ${ [-a,a] }$,
where ${ a \!=\! 3L/2 }$.
Using equation~\eqref{eq:final_Eddington}, we have
\begin{subequations}
\begin{align}
\rho(x) &= \tfrac{1}{2} \frac{M}{a} \Theta(a \!-\! |x|) ,
\label{eq:def_rho_harmonic}
\\
\psi(x)&=
\begin{cases}
E_{a} \, |x / a| & \mathrm{if} \, |x| \geq a,
\label{eq:def_psi_harmonic}
\\
\displaystyle{ \tfrac{1}{2} E_{a} + \tfrac{1}{2} \omega_a^2 x^2 }& \mathrm{if} \, |x|< a ,
\end{cases}
\\
F(E) & = \frac{M}{2\pi a} \frac{1}{\sqrt{2(G M a \!-\! E)}},
\label{eq:df_harmonic}
\end{align}
\label{eq:all_definitions_harmonic}\end{subequations}
where we introduced
${ E_{a} \!=\! G M a }$ and 
${ \omega_a \!=\! \sqrt{GM/a} }$. 
The total energy is ${ \Etot \!=\! \frac{1}{2} G M^2 a \!=\! \frac{3}{4} G M^2 L }$. 
As expected, \scHarmonic{} has a quadratic potential
in its central region.
As such, this potential is degenerate:
all particles share the same orbital frequency.
Indeed, all particles satisfy ${ x_{a} \!\leq\! a }$,
so that their individual energy satisfies
${ E_{a} / 2 \!\leq\! E \!\leq\! E_{a} }$.
Following equation~\eqref{eq:Omega},
we find
\begin{align}
\frac{1}{\Omega} = \frac{2}{\pi} \sqrt{\frac{a}{GM}} \int_{0}^{x_{\ra}}\hspace*{-1mm} \frac{\rd x}{\sqrt{x_{\ra}^2 \!-\! x^2}} = \sqrt{\frac{a}{GM}},
\end{align}
As expected, all particles have the same initial mean-field orbital frequency ${ \Omega \!=\! \omega_a }$.

\subsection{\scAnharmonic{}}
\label{app:anharmonic_distribution}

We now focus on the \scAnharmonic{} potentials
which are compact but partially degenerate (see Table~\ref{tab:Potentials}).
Given ${ 0 \!<\! \eps \!<\! 1 }$,
we start from the density profile
${ \rho_{\eps} \!=\! \rhoH \ast \Theta (\eps a \!-\! |x|) }$,
where ${ \rhoH }$ is the harmonic density (equation~\ref{eq:def_rho_harmonic}),
${ \ast }$ is the convolution operator,
and ${ \Theta }$ the usual Heaviside function.
In practice, one finds that the density profile reads
\begin{align}
\rho_{\eps}(x) & = \frac{M}{2a}
\begin{cases}
\displaystyle 1 & \mathrm{if} \, |x/a| \leq 1 - \eps ,
\\
\displaystyle 0 & \mathrm{if} \, |x/a| \geq 1 + \eps ,
\\
\displaystyle{\frac{1 \!+\! \eps \!-\! |x/a|}{2\eps } } & \mathrm{otherwise}.
\end{cases}
\end{align}
The associated potential is (equation~\ref{eq:def_psi_i})
\begin{align}
\psi_{\eps}(x) &\!=\! G M a
\begin{cases}
\frac{1}{2} \!+\! \frac{\eps^2}{6} \!+\! \frac{(x/a)^2}{2}& \!\!\mathrm{if} \, |x/a| \!\leq\! 1 \!-\! \eps ,
\\
|x / a|&\!\! \mathrm{if} \, |x/a| \!\geq\! 1 \!+ \!\eps ,
\\ 
\frac{1 + 3 (\eps -|x/a|) + 3 (\eps +|x/a|)^2+ (\eps -|x/a|)^3}{12 \eps }&\! \! \mathrm{otherwise}.
\end{cases}
\end{align}
We can also compute the associated DF.
From equation~\eqref{eq:def_rho_anharmonic},
we find that the density derivative is
\begin{align}
\rho'_{\eps}(x) &= - \frac{M}{ 4a^2 \eps}
\begin{cases}
\sgn(x/a)& \mathrm{if} \,1 \!-\! \eps \leq |x/a|\leq 1 \!+\! \eps ,
\\
\hspace*{5mm} 0& \mathrm{otherwise}.
\end{cases}
\end{align}
We let ${ \Psi_{\pm} \!=\! \psi_{\eps} [a(1 \!\pm\! \eps)] }$,
and introduce ${ E_{\min} \!=\! \psi_{\eps}(0) \!>\! 0 }$.
Using equation~\eqref{eq:final_Eddington}, we find
\begin{align}
F(E) &= \frac{M}{4\pi\sqrt{2} a^2 \eps}
\begin{cases}
\displaystyle 0 & \mathrm{if} \, \Psi_{+} \!<\! E ,
\\
\displaystyle \int_{x_{\ra}}^{a(1+\eps)} \hspace*{-4mm} \frac{\rd x}{ \sqrt{\psi(x) \!-\! E}} & \mathrm{if} \, \Psi_{-} \!\leq\! E \!<\! \Psi_{+} ,
\\[2.0ex]
\displaystyle \int_{a(1-\eps)}^{a(1+\eps)} \hspace*{-3mm} \frac{\rd x}{ \sqrt{\psi(x) \!-\! E}} & \mathrm{if} \, E_{\min} \!\leq\! E \!<\! \Psi_{-} .
\end{cases}
\label{eq:final_DF_anharmonic}
\end{align}
In the limit ${ \eps \!\rightarrow\! 0^{+} }$,
the latter expression becomes a finite difference, to give
\begin{equation}
F(E) = \frac{M}{2\pi\sqrt{2} a } \frac{1}{ \sqrt{\psi(a) \!-\! E}} = \frac{M}{2\pi a } \frac{1}{ \sqrt{2(GMa \!-\! E)}} .
\end{equation}
This is exactly the harmonic DF from equation~\eqref{eq:df_harmonic}.
Using equation~\eqref{eq:calc_Etot},
we find that the total energy is
\begin{align}
E_{\rtot , \eps} = \tfrac{1}{2} G M^2 a \,\big(1 \!+\! \tfrac{1}{10} \, (5 \!-\! \eps) \, \eps^{2} \big).
\end{align}

\begin{figure}
\centering
\includegraphics[width=0.45 \textwidth]{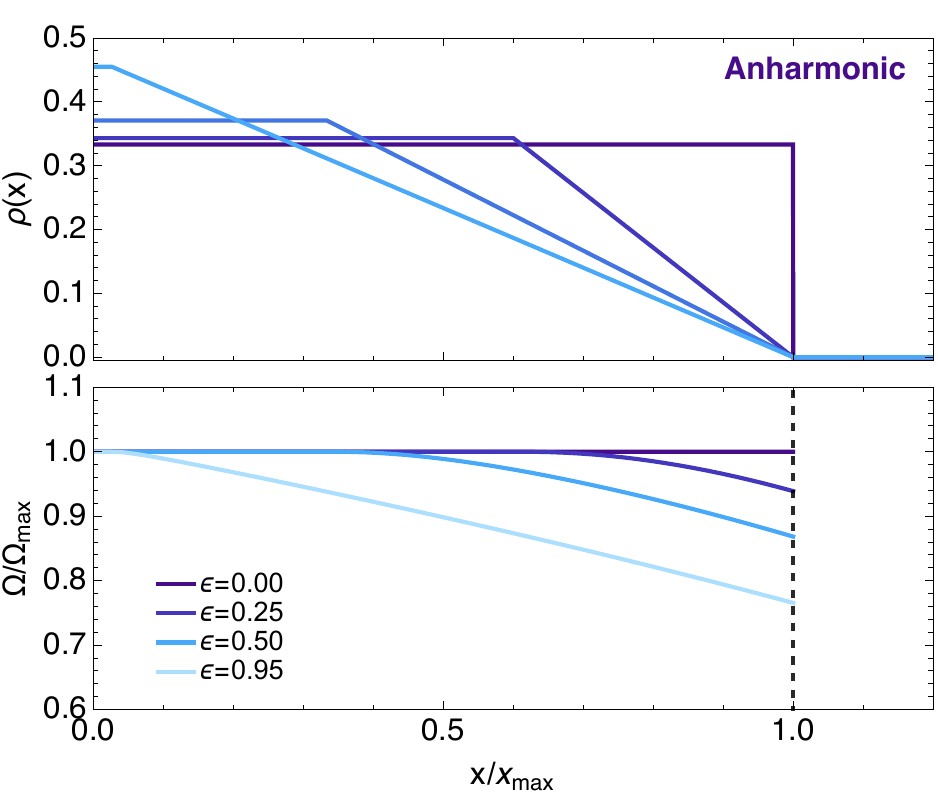}
\caption{Illustration of the \scAnharmonic{} family
for various values of $\eps$.
We define $\Omega_{\max}$ and $x_{\max}$ as in Figure~\ref{fig:Illustration_Frequency}. 
These systems are of finite radial support,
with a central constant density core,
i.e.\ a region that is dynamically degenerate.
}
\label{fig:Illustration_Anharmonic}
\end{figure}

To ease the comparison between different values of $\eps$,
we want the total energy not to depend on $\eps$.
To that end, we consider some (renormalized) pairs of potentials and densities
\begin{subequations}
\begin{align}
\rho_{\eps,A}(x) & = A \, \rho_{\eps}(A x),
\label{eq:def_rhoA}
\\
\psi_{\eps,A}(x) & = \psi_{\eps}(A x) / A ,
\label{eq:def_psiA}
\end{align}
\label{eq:def_rhopsiA}\end{subequations}
whose total mass is $M$.
The associated renormalized DF is simply
\begin{align}
F_{A}(x_{\ra}) = A^{3/2} \,F(A \, x_{\ra}) ,
\end{align}
and its total energy is ${ E_{\rtot, \eps, A} \!=\! E_{\rtot , \eps} / A }$. 
Therefore, if we let
\begin{equation}
A = A(\eps) = 1 \!+\! \tfrac{1}{10} (5 \!-\! \eps) \eps^2 ,
\label{eq:val_A_anharmonic}
\end{equation}
we find ${ E_{\rtot, \eps,A[\eps]} \!=\! \frac{1}{2} G M^2 a }$.
Imposing ${ a \!=\! 3 L / 2 }$,
we ensure therefore that all \scAnharmonic{} potentials,
independently of their values of $\eps$,
have the same total energy, ${ \Etot \!=\! \frac{3}{4} G M^{2} L }$.
It is this renormalised family of potentials
that is used in the main text.
In figure~\ref{fig:Illustration_Anharmonic},
we illustrate the \scAnharmonic{} clusters
for various values of $\eps$.

For a given value of $\eps$,
the edge of the constant density core is ${ |x/a| \!=\! (1 \!-\! \eps) / A (\eps) }$.
This edge goes to 1 for ${ \eps \!\to\! 0 }$,
i.e.\ the fully degenerate harmonic case,
and to 0 for ${ \eps \!\to\! 1 }$,
i.e.\ the fully non-degenerate case.
The fraction of particles in the constant density core
is ${ 1 \!-\! \eps }$.
Therein, the (constant) orbital frequency (equation~\ref{eq:Omega}) is
${ \Omega \!=\! \sqrt{A(\eps)} \, \omega_a }$,
with $\omega_a$ introduced in equation~\eqref{eq:def_psi_harmonic}.

\section{Balescu--Lenard equation}
\label{app:BL}

\subsection{1D case}
\label{app:BL_1D}

Following equation~{(7)} of~\cite{Roule+2022},
the inhomogeneous Balescu--Lenard equation reads
\begin{align}
\frac{\p F(J , t)}{\p t} = {} &  2 \pi^{2} m \frac{\p }{\p J} \bigg[ \sum_{k , k'} k \!\! \int \!\! \rd J' \, |U_{k k'}^{\rd} (J , J' ; k \Omega [J]) |^{2}
\label{eq:BL_app}
\\
\times {} & \, \deltaD (k \Omega [J] \!-\! k' \Omega [J'] )  \bigg( k \frac{\p }{\p J} - k' \frac{\p }{\p J'} \bigg) F(J , t) F (J' , t) \bigg] ,
\nonumber
\end{align}
with $k,k'$ the resonance numbers,
${ J \!=\! \!\oint\! \rd x \, v / 2 \pi }$ the action~\citep[see, e.g.\@, appendix~{A.1} in][]{Roule+2022},
and ${ \Omega (J) }$ the orbital frequency.
The system's DF, ${ F (J , t) }$, is normalised so that ${ \!\int\! \rd \theta \rd J \, F (J, t) \!=\! M }$, with $\theta$ the angle coordinate associated with $J$.
In the main text, for the sake of clarity,
we presented \BL\@ in equation~\eqref{eq:BL}
using the energy, $E$, as the representative coordinate for the orbits.
In 1D, such a rewriting can be directly obtained from equation~\eqref{eq:BL_app}
using the relation ${ \rd E / \rd J \!=\! \Omega (J) }$.

Equation~\eqref{eq:BL_app} also involves the frequency-dependent
dressed coupling coefficients,
${ U^{\rd}_{kk'} (J , J' ; \omega) }$.
These coefficients capture the effect of the self-gravitating amplification.
Generically, they can be obtained from a self-consistent 
Dyson-like equation reading~\citep{Luciani+1987}
\begin{align}
U^{\rd}_{kk'} (J,J'; {} & \omega)  = U_{kk'} (J,J') 
\label{eq:Dyson}
\\
+ {} &  2 \pi \sum_{k''} \!\! \int \!\! \rd J'' \, \frac{k'' \p F / \p J''}{k'' \Omega [J''] - \omega} U_{k k''} (J , J'') \, U^{\rd}_{k''k'} (J'' , J' ; \omega) .
\nonumber
\end{align}
In that expression, we introduced the bare coupling coefficients,
${ U_{kk'} (J , J') }$. These are given by the Fourier transform
in angles of the pairwise interaction, ${ U (x , x') }$, from equation~\eqref{eq:def_U_1d}.
Following~\cite{Roule+2022}, these coefficients read
\begin{equation}
U_{kk'} (J , J') = \frac{1}{2\pi} \!\! \int \!\! \rd \theta \, \rd \theta' \, U [x(\theta,J) , x'(\theta',J')] \, \re^{- \ri (k \theta - k' \theta')} .
\label{eq:def_Uk}
\end{equation}
In practice, the self-consistent definition from equation~\eqref{eq:Dyson}
can be explicitly inverted using the basis method from~\cite{Kalnajs1976II}.
Its tailoring to the case of 1D gravity is detailed in appendix~{A.2} of~\cite{Roule+2022}.

The Balescu--Lenard equation (equation~\ref{eq:BL_app})
is the master quasilinear kinetic equation
describing the mean long-term relaxation of an integrable isolated
finite-$N$ self-gravitating system.
In practice, equation~\eqref{eq:BL_app} can be derived through a couple of venues,
including the BBGKY hierarchy~\citep{Heyvaerts2010},
the Klimontovich equation~\citep{Chavanis2012},
the Novikov theorem~\citep{Fouvry+2018} and
the Rostoker principle~\citep{Hamilton2021}.
In essence, all these approaches are based on a two-timescale
approach that: (i) solves, at leading order,
for the dynamics of fluctuations
on the (fast) dynamical time;
(ii) determines the impact of the quadratic coupling
of these flucutations on the (slow) relaxation time.
We refer to~\cite{Hamilton+2024} for a detailed review
of these various works.

Yet, \BL\ holds assuming a few key hypotheses.
In particular, it assumes that: 
(i) the mean potential is integrable
so that global angle-action coordinates exist;
(ii) the system is only subject to internal finite-$N$ perturbations;
(iii) fluctuations must be small so that they may treated perturbatively;
(iv) the collective amplification through self-gravity
must remain limited, so that linear response theory applies;
(v) it solely focuses on the mean relaxation
by performing an ensemble-average over independent
realisations of the initial conditions;
(vi) the system's frequency profile must remain non-degenerate,
i.e.\ phase mixing must always be active,
so that long-term resonances are meaningful.
For the case considered in the present work,
this last hypothesis does not hold anymore  for \scHarmonic{}.
Indeed, 
all the orbits then share the exact same orbital frequency.
As a consequence, at linear order (which \BL\@ assumes),
fluctuations always remain in phase.
This makes their long-term quasilinear interaction intrinsically ill-posed.
The present work aims at exploring, numerically,
the signatures of the relaxation
in \scHarmonic{},
where one the underlying hypothesis of \BL\ is violated.

\subsection{3D case}
\label{app:BL_3D}

In 3D, for a spherically symmetric system (hence integrable),
the \BL\ equation takes the form~\citep[see, e.g.\@,][]{Hamilton+2018,Fouvry+2021}
\begin{align}
\frac{\p \oF (\obJ , t)}{\p t} = {} & \pi (2\pi)^{3} m \frac{\p }{\p \obJ} \!\cdot\! \bigg[ \sum_{\obk,\obkp} \obk \!\! \int \!\! \rd \obJp \, |\oU^{\rd}_{\obk\obkp} (\obJ , \obJp ; \obk \!\cdot\! \obO [\obJ] )|^{2}
\label{eq:BL_3D_full}
\\
\times {} & \deltaD (\obk \!\cdot\! \obO [\obJ] \!-\! \obkp \!\cdot\! \obO [\obJp]) \, \bigg( \obk \!\cdot\! \frac{\p }{\p \obJ} \!-\! \obkp \!\cdot\! \frac{\p }{\p \obJp} \bigg) \oF (\obJ , t) \, \oF (\obJ , t) 
\bigg],
\nonumber
\end{align}
with ${ \obk \!\in\! \mathbb{Z}^{3} }$ 
the resonance vector,
$\obJ$ the vector of actions
and ${ \obO [\obJ] }$ the associated vector of orbital frequencies.
We also introduced ${ \oF (\obJ , t) }$
as the system's DF in action space,
normalised so that ${ \!\int\! \rd \obJ \rd \obT F \!=\! M }$,
with $\obT$ the vector of angles.
We refer to appendix~{D.1} of~\cite{Fouvry+2021}
for a precise definition
of the 3D dressed coupling coefficients, ${ \oU^{\rd}_{\obk,\obkp} (\obJ , \obJp ; \omega) }$.
In the coming paragraphs, we detail how equation~\eqref{eq:BL_3D_full}
also becomes ill-defined in the case of a 3D harmonic system,
just like equation~\eqref{eq:BL} in 1D.

In a 3D spherically symmetric system,
a typical choice is to pick the action coordinates to be
${ \obJ \!=\! (J_{r} , L , L_{z}) }$,
with $J_{r}$ the radial action,
$L$ the norm of the angular momentum vector
and $L_{z}$ its projection along some given $z$-axis~\cite{Binney2008}.
Then, for a spherically symmetric system,
two additional symmetries arise:
(i) the system's DF has the dependency
${ F \!=\! F(\obJ , t) \!=\! F (J_{r} , L , t) }$;
(ii) the associated orbital frequency
are given by ${ \obO \!=\! (\Omega_{r} , \Omega_{\phi} , 0) }$
involving respectively the radial and azimuthal frequencies.
Importantly, because the orbital plane is conserved by the mean field dynamics,
the third frequency, $\Omega_{z}$, vanishes.

As a result of these two symmetries,
as first shown in~\cite{Hamilton+2018},
it is possible to ``integrate'' equation~\eqref{eq:BL_3D_full} over $L_{z}$,
The \BL\ equation then becomes 
\begin{align}
\frac{\p F (\bJ , t)}{\p t} {} & \propto \frac{\p }{\p \bJ} \!\cdot\! \bigg[ \sum_{\bk,\bkp} \bk \!\! \int \!\! \rd \bJp \, |U^{\rd}_{\bk\bkp} (\bJ , \bJp ; \bk \!\cdot\! \bO [\bJ] )|^{2}
\label{eq:BL_3D_to_2D}
\\
\times {} & \deltaD (\bk \!\cdot\! \bO [\bJ] \!-\! \bkp \!\cdot\! \bO [\bJp]) \, \bigg( \bk \!\cdot\! \frac{\p }{\p \bJ} \!-\! \bkp \!\cdot\! \frac{\p }{\p \bJp} \bigg) F (\bJ , t) \, F (\bJ , t) .
\bigg]
\nonumber
\end{align}
In that expression, ${ \bk \!=\! (k_{r} , k_{\phi}) \!\in\! \mathbb{Z}^{2} }$
is the ``in-plane'' resonance vector,
${ \bJ \!=\! (J_{r} , L) }$ the in-plane actions,
and ${ \bO \!=\! (\Omega_{r} , \Omega_{\phi}) }$
the in-plane frequencies.
We also introduced the ``reduced'' DF,
${ F (\bJ) \!=\! 2 L \oF (\bJ) }$.
We refer to appendix~{D.2} in~\cite{Fouvry+2021}
for the derivation of equation~\eqref{eq:BL_3D_to_2D}
along with the detailed expression
of the in-plane (dressed) coupling coefficients,
${ U^{\rd}_{\bk\bkp} (\bJ , \bJp ; \omega) }$.
At this stage, the crucial point is to note
that equation~\eqref{eq:BL_3D_to_2D}
has now become an effectively two-dimensional diffusion equation.

Armed with equation~\eqref{eq:BL_3D_to_2D},
we can now consider the case of a 3D spherically symmetric harmonic system.
In that case, the unperturbed orbits are closed ellipses
that are all skimmed with the same frequency $\omega_{0}$.
Because the cluster's centre lies at the centre of the orbital ellipses,
the in-plane frequencies, ${ \bO (\bJ) }$,
take the simple form
\begin{equation}
\bO (\bJ) = (2 \, \omega_{0} , \omega_{0}) .
\label{eq:harmonic_frequencies_3D}
\end{equation}
Importantly, we note that ${ \bO (\bJ) }$ is completely independent of $\bJ$,
i.e. independent of the considered orbit.
Equation~\eqref{eq:harmonic_frequencies_3D} is the 3D equivalent
of the flat frequency profile,
${ \Omega (E) \!=\! \cst }$ considered in 1D for \scHarmonic{}.
With such a frequency profile,
the resonance condition from equation~\eqref{eq:BL_3D_to_2D} simply 
becomes
\begin{equation}
\deltaD (\bk \!\cdot\! \bO [\bJ] \!-\! \bkp \!\cdot\! \bO [\bJp]) = \frac{1}{\omega_{0}} \, \deltaD (2 [k_{r} - k^{\prime}_{r}] + [k_{\phi} - k^{\prime}_{\phi}]) .
\label{eq:resonance_condition_3D}
\end{equation}
Because the argument of this function is independent of $\bJ$ and $\bJp$,
this makes equation~\eqref{eq:BL_3D_to_2D} completely ill-defined.
In particular, local resonances,
i.e.\ the choice ${ \bk \!=\! \bkp }$,
leads to a systematic ${ \deltaD (0) }$ in equation~\eqref{eq:BL_3D_to_2D}.
This is mathematically meaningless.
As a conclusion, 
quasilinear kinetic theories like \BL\@
are also unable to describe the self-consistent relaxation of harmonic systems in 3D. 
Given that the 1D and 3D harmonic suffer from the
exact same ``over-abundance'' of resonances,
this justifies our focus, in the main text,
on the characterisation of the relaxation of 1D harmonic systems:
these are much easier to simulate on long timescales
compared to their 3D equivalents.

In practice, the long-term relaxation of 3D harmonic spheres
was considered in detail by~\cite{Sellwood2015}, hereafter~\citetalias{Sellwood2015}.
In practice, \citetalias{Sellwood2015} used various types of simulation codes
(tree code, spherical grid, polar grid, and basis field expansion)
for that numerical exploration.
Focusing on \citetalias{Sellwood2015}'s result
for the harmonic case, it showed that:
(i) the diffusion of individual particles in harmonic systems
is accerated compared to their non-degenerate equivalents~\citepalias[see figure~{4} in][]{Sellwood2015}.
Indeed, the typical diffusion time of test particles
was found to scale like $N^{-1/2}$
compared to the (typically expected) scaling in $N^{-1}$
observed in other non-degenerate systems.
Interestingly this is the same scaling that was shown
in~\cite{DiCintio+2025}
to play an important role regarding dynamical friction
in harmonic systems.
(ii) Looking at the overall relaxation of harmonic clusters,
\citetalias{Sellwood2015} offered some hints that its relaxation
was greatly delayed compared to the other clusters~\citepalias[see figure~{1} in][]{Sellwood2015}. We argue that this (extremely) slow relaxation
hinted by \citetalias{Sellwood2015}
is the 3D equivalent of the 1D delayed relaxation
that we put forward in Figure~\ref{fig:scaling_plummer_harmonic}
using very long-term simulations.
Unfortunately, \citetalias{Sellwood2015}
did not perform any precise measurements
of the dependence of the cluster's relaxation rate
as a function of the number of particles.
This will be the topic of future numerical explorations.

\section{Thermodynamical equilibria}
\label{app:Thermodynamical_distribution}

In 1D, self-gravitating systems admit genuine thermodynamical equilibria,
i.e.\ the statistical distribution
toward which any $N$-body realisation unavoidably relaxes.
It is this distribution that is used in Figure~\ref{fig:ks}
to track long-term relaxation.

Since we are performing simulations with rather small values
of $N$, it is important to obtain the expression of these equilibria
while accounting exactly for finite-$N$ effects.
Fortunately, this calculation was performed explicitly in~\cite{Rybicki1971},
whose main results we now reproduce.
Let us assume that we are given a realisation of an $N$-body system
with $N$ the total number of particles,
$\Etot$ the total energy (equation~\ref{eq:def_Etot}),
and a vanishing total linear momentum (${ \Ptot \!=\! \sum_{i} m_{i} v_{i} }$).\footnote{This can always be assumed through a change of inertial frame.}

Following~\citet{Rybicki1971}, the exact density profile
of the $N$-particle thermodynamical equilibrium distribution reads
\begin{equation}
\rhoth (x) = \frac{2M}{L}\bigg(1\!-\! \frac{5}{3N}\bigg) \sum_{\ell=1}^{N-1} \,A_{\ell}^{(N)} \,\bigg(1 \!-\! \frac{4\, \ell\, |x/L|}{3 N}\bigg)_{+}^{\frac{3}{2}N-\frac{7}{2}},
\label{eq:def_rho_thermodynamical_N}
\end{equation}
where we define ${ (u)_{+} \!=\! \max(u,0) }$ and
\begin{align}
A_{\ell}^{(N)} &= \frac{[(N \!-\! 1)!]^2\, (-1)^{\ell-1}\ell}{(N \!-\! 1 \!-\! \ell)!\,(N \!-\!1 \!+\! \ell)!}.
\end{align}
In the limit ${ N \!\to\! + \infty }$, equation~\eqref{eq:def_rho_thermodynamical_N}
becomes the (collisionless) thermodynamical equilibrium distribution~\citep[see, e.g.\@,][and references therein]{Joyce+2010}
\begin{equation}
\label{eq:def_rho_thermodynamical_inf}
\rhoth (x) = \frac{M}{2} \sech^2 (x/L).
\end{equation}
From equation~\eqref{eq:def_rho_thermodynamical_N},
we find that the associated cumulative mass distribution,
${ \Mth (x) \!=\! \int_{- \infty}^{x} \rd y \rho(y) }$, reads
\begin{align}
\Mth (\leq x) {} & \! = \frac{M}{2} \bigg(1 \!-\! \frac{5}{3N} \!\bigg) \sum_{\ell=1}^{N-1} \frac{3 N A_{\ell}^{(N)}}{\ell} \int_{-\infty}^{4\, \ell\, x/(3 L N)} \hspace*{-12mm} \rd z\,\big(1 \!-\! |z|\big)_{+}^{\frac{3}{2}N-\frac{7}{2}}
\nonumber
\\
{} & \! = \frac{M}{2} \bigg(1 \!-\! \frac{5}{3N} \!\bigg) \! \sum_{\ell=1}^{N-1} \! \frac{3 N A_{\ell}^{(N)}}{\ell} \mI \big[ \tfrac{3}{2}N \!-\! \tfrac{7}{2},4\, \ell\, x/(3 L N) \big] .
\end{align}
where, for ${ \beta \!>\! 0 }$, we introduced
\begin{align}
\mI [\beta,w]
& = \int_{-\infty}^{w} \hspace*{-2mm}\rd z\,\big(1 \!-\! |z|\big)_{+}^{\beta} 
\notag
\\
& =\frac{1}{\beta \!+\! 1}
\begin{cases}
\big(1 \!-\! |w|\big)_{+}^{\beta+1}& \mathrm{if} \hspace*{2mm} w < 0,
\\
2 \!-\! \big(1 \!-\! |w| \big)_{+}^{\beta+1}& \mathrm{if} \hspace*{2mm} w \geq 0.
\end{cases}
\end{align}

In practice, the calculation of $A_{\ell}^N$ in equation~\eqref{eq:def_rho_thermodynamical_N}
can be accelerated through the recursion
\begin{equation}
A_{1}^{(N)} =\frac{N \!-\! 1}{N} \quad;\quad
A_{\ell+1}^{(N)}=\frac{(\ell \!+\! 1 \!-\! N) (\ell \!+\! 1)}{\ell\,(\ell \!+\! N)}\,A_{\ell}^{(N)} .
\end{equation}
Since ${ (\ell \!+\! 1 \!-\! N) / \ell \!\leq\! 1 }$
and ${ (\ell \!+\! 1) / ( \ell \!+\! N) \!\leq\! 1 }$
for ${ N \!\geq\! 1 }$, this process is numerically stable.

\section{Generation of initial conditions}
\label{app:ic_generation}

In this Appendix,
we detail our approach to sample the initial conditions
of the $N$-body realisations.
Our goal is to sample $N$ particles ${ \{(x_i,v_i)\}_i }$
for a system with DF ${ F(E) }$, potential ${ \psi(x) }$ and density ${ \rho(x) }$.
We can achieve this through two methods.
For the sake of completeness, we describe them briefly below.

\subsection{Inverse transform sampling}
\label{app:inv_sampling}

We consider the spatial cumulative distribution function
\begin{align}
\CDF_x(x)&=\frac{1}{M} \int_{-\infty}^x \hspace{-2mm}\rd x' \rho(x'),
\end{align}
and the conditional velocity cumulative distribution function
\begin{align}
\CDF_{v|x}(v|x)&=\frac{1}{\rho(x)}\int_{-\infty}^v \hspace{-2mm}\rd v' F(E[x,v']),
\end{align}
where ${ E[x,v] \!=\! \psi(x) \!+\! v^2 / 2 }$.
We assume that their inverse functions ${ \CDF^{-1}_x (u_x) }$
and ${ \CDF^{-1}_{v | x} (u_v | x ) }$ can be computed either analytically or numerically.
Then, we can sample each position-velocity pairs via the inverse transform method~\citep{devroye1986}.
It proceeds as follows:
\begin{enumerate}
\item We draw ${ u_x \!\in\! \mU (0,1) }$, from the uniform distribution.
\item We set ${ x \!=\! \CDF^{-1} (u_x) }$.
\item We draw ${ u_v \!\in\! \mU(0,1) }$.
\item We set ${ v \!=\! \CDF^{-1}_{v | x}(u_v | x) }$.
\end{enumerate}
In practice, we use this approach for \scPlummer{} and \scHarmonic{}. 
Indeed, for \scPlummer{},
we find from equation~\eqref{eq:def_Plummer} that
\begin{subequations}
\begin{align}
\CDF_x (x) & = \frac{1}{2}\bigg(1 + \frac{x}{\sqrt{x^2+\alpha^2}}\bigg),
\\
\CDF_{v|x} (v|x)& = \frac{15 G^3 M^4 \alpha^2}{32 \rho(x)\psi(x)^3} \frac{15s\!+\!20s^3\!+\!8s^5\!+\!8(1\!+\!s^2)^{5/2}}{15(1\!+\!s^2)^{5/2}},
\end{align}
\end{subequations}
where ${ s \!=\! v / \sqrt{2 \psi(x)} }$. The inverse for the spatial CDF reads
\begin{align}
\CDF^{-1}_x (u_x) & = \frac{\alpha \, (2u_x \!-\! 1)}{\sqrt{1 \!-\! (2u_x \!-\! 1)^2}},
\end{align}
whereas the inverse for the velocity CDF can be computed numerically using bisection.

For \scHarmonic{}, the spatial sampling reduces to sampling $x$ uniformly over ${ [-a,a] }$.
Then, the velocity CDF is given by
\begin{align}
\CDF_{v|x} (v|x) & = \tfrac{1}{2} + \frac{\arcsin(v/v_{\max}[x]) }{\pi},
\end{align}
where ${ v_{\max}^2[x] \!=\! E_a \!-\! \omega_a^2 x^2 }$.
Its inverse reads
\begin{align}
\CDF^{-1}_{v|x} (u_v|x) & = - v_{\max}[x] \cos (\pi u_v).
\end{align}

\subsection{Rejection sampling}
\label{app:rjt_sampling}

We now suppose that the inverse of the CDF are not available,
as is the case for \scCompact{} and \scAnharmonic{}.
For these systems, we use the von Neumann rejection method~\citep{Chib1995,Robert2004}.
It proceeds as follows.

We wish to sample a random variable $\mR$ described by the density ${ \pi(r) \!=\! K f(r) }$, whose normalisation constant ${ K \!>\! 0 }$ is unknown.
Suppose that $\mR$ takes values over a finite interval of size $L$. We consider the uniform distribution $\mZ$ over that interval and the uniform distribution $\mU$ over ${ [0,1] }$. We let $\mM$ be an upper bound of ${ f(r) }$ over its interval of definition. Then, we can obtain a sample from ${ \pi(r) }$ through the following algorithm:
\begin{enumerate}
\item We generate a proposal, $z$, from $\mZ$ and a value $u$ from $\mU$.
\item If ${ \mM \, u \!\leq\! f(z) }$, then we accept the proposal $z$.
\item If not, we loop back to step 1.
\end{enumerate}
We can now apply this algorithm to generate a pair of position-velocity ${ (x,v) }$.
First, to sample the position, we apply the rejection algorithm to ${ r\!\leftarrow\! x }$, ${ f\!\leftarrow\! \rho }$ and ${ \mM \!\leftarrow\! \max_{x} \rho(x) }$.
Then, to sample the velocity,
we apply the algorithm to ${ r\!\leftarrow\! v|x }$, ${ f\!\leftarrow\! F }$ and ${\mM \!\leftarrow\! \max_{E} F(E) }$,
where $x$ is the position we just sampled.

\section{Complementary figures}
\label{app:ks_val_additional}

\begin{figure}
\centering
\includegraphics[width=0.45 \textwidth]{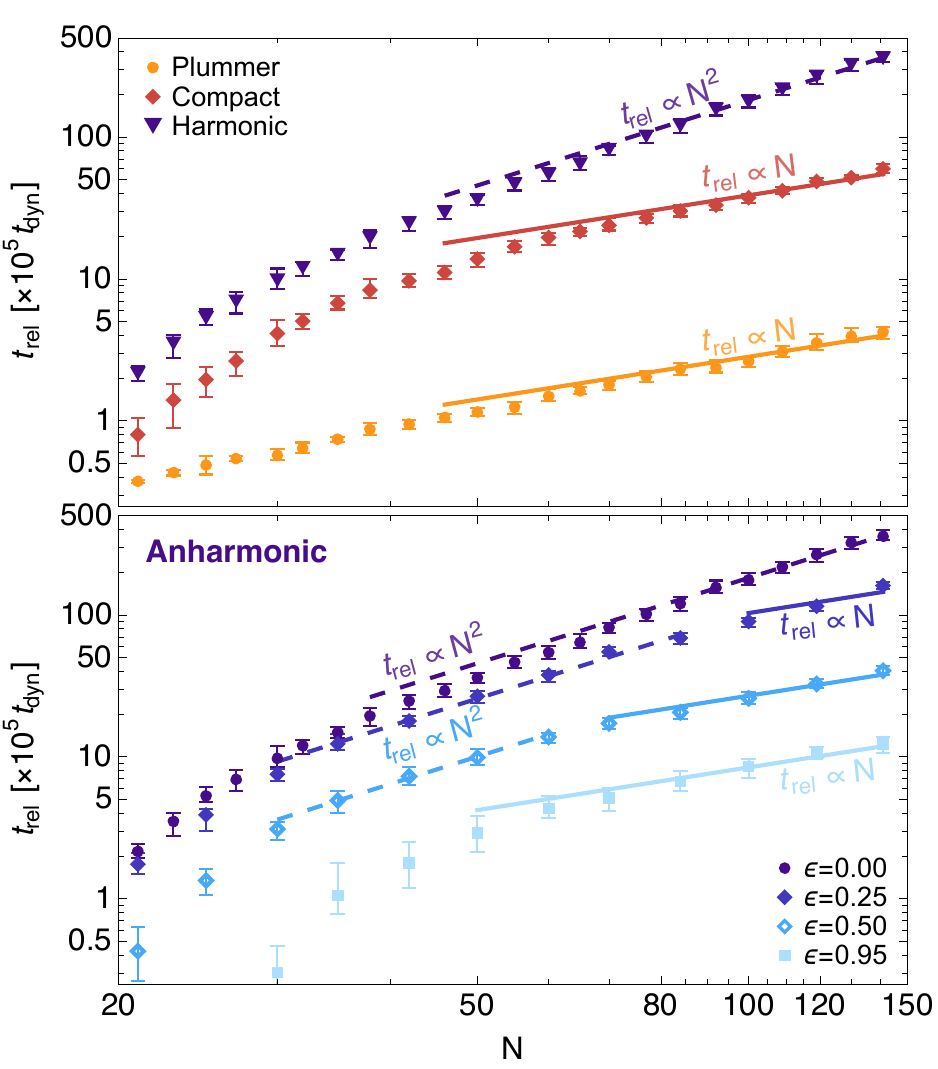}
\caption{Same as Figure~\ref{fig:scaling_plummer_harmonic} (top panel)
and Figure~\ref{fig:scaling_anharmonic} (bottom panel),
here using the threshold ${ D_0 \!=\! 0.022 }$ (equation~\ref{eq:introduction_D0}).
We recover the exact same trends as in the main text.
}
\label{fig:scaling_app}
\end{figure}

In this Appendix, we complement Figures~\ref{fig:scaling_plummer_harmonic}
and~\ref{fig:scaling_anharmonic},
by considering a different threshold, $D_{0}$,
for the measurements of the relaxation time (as introduced in equation~\ref{eq:introduction_D0}).
In Figure~\ref{fig:scaling_app},
we consider the value ${ D_{0} \!=\! 0.022 }$.
As visible in Figure~\ref{fig:ks},
this corresponds to an earlier measurement
compared to Figures~\ref{fig:scaling_plummer_harmonic}
and~\ref{fig:scaling_anharmonic}.
Reassuringly, we recover the exact same trends
as in the main text, namely
(i) the thermodynamic blocking of \scHarmonic{}
which relaxes on a timescale of order ${ \mO (N^{2}) }$;
(ii) the transition between the two regimes of relaxation
for \scAnharmonic{}, as one varies $\eps$,
the level of dynamical degeneracy.

\end{appendix}
\end{document}